\documentclass[fleqn,10pt]{wlscirep}
\usepackage[utf8]{inputenc}
\usepackage[T1]{fontenc}

\title{Robust statistical properties of T1 transitions in confluent cell tissues}

\author[1,*]{Harish P Jain}
\author[2,3,4]{Axel Voigt}
\author[1]{Luiza Angheluta}
\affil[1]{Njord Centre, Department of Physics, University of Oslo, Oslo, 0371, Norway}
\affil[2]{Institute of Scientific Computing, Technische Universität Dresden, Dresden, 01062, Germany}
\affil[3]{Center of Systems Biology Dresden, Pfotenhauerstr. 108, 01307 Dresden, Germany}
\affil[4]{Cluster of Excellence - Physics of Life, TU Dresden, 01062 Dresden, Germany}

\affil[*]{harishpj@fys.uio.no}

\begin{abstract}
Large-scale tissue deformation which is fundamental to tissue development hinges on local cellular rearrangements, such as T1 transitions. In the realm of the multi-phase field model, we analyse the statistical and dynamical properties of T1 transitions in a confluent tissue. We identify an energy profile that is robust to changes in several model parameters. It is characterized by an asymmetric profile with a fast increase in energy before the T1 transition and a sudden drop after the T1 transition, followed by a slow relaxation. The latter being a signature of the fluidity of the cell tissue. 
We show that T1 transitions are sources of localised large deformation of the cells undergoing the neighbour exchange and induce other T1 transitions in the nearby cells through a chaining of events that propagate local cell deformation to large scale tissue flows. 
\end{abstract}
\begin{document}

\flushbottom
\maketitle
\thispagestyle{empty}

\section{Introduction}

Collective motion of cells is essential to several processes including development of an embryo, tissue morphogenesis, wound healing, homeostasis and cancer metastasis~\cite{Ladoux2017,bruguesForcesDrivingEpithelial2014,friedlClassifyingCollectiveCancer2012}. These biological processes are highly complex and orchestrate mechanical, chemical and biochemical interactions across multiple scales~\cite{guiraoUnifiedQuantitativeCharacterization2015,etournayInterplayCellDynamics2015,iyerEpithelialViscoelasticityRegulated2019,dyeSelforganizedPatterningCell2021}. Through the interplay between directed motion, neighbour alignment and mechanical interactions, cell tissues exhibit emergent structures and dynamics that are crucial for their biological function. A fundamental underlying process for emergent large-scale behavior is the topological rearrangement of  neighbouring cells, also known as T1 transition. It is a local, dissipative event that leads to remodelling of the tissue architecture and influences the large-scale flow properties of cell tissues that affects tissue homeostasis and epithelial morphogenesis~\cite{kellerMechanismsConvergenceExtension2000, walck-shannonCellIntercalationTop2014, rauziCellIntercalationSimple2020}.

\begin{figure}[t!]
    \centering
    \includegraphics[width=0.9\textwidth]{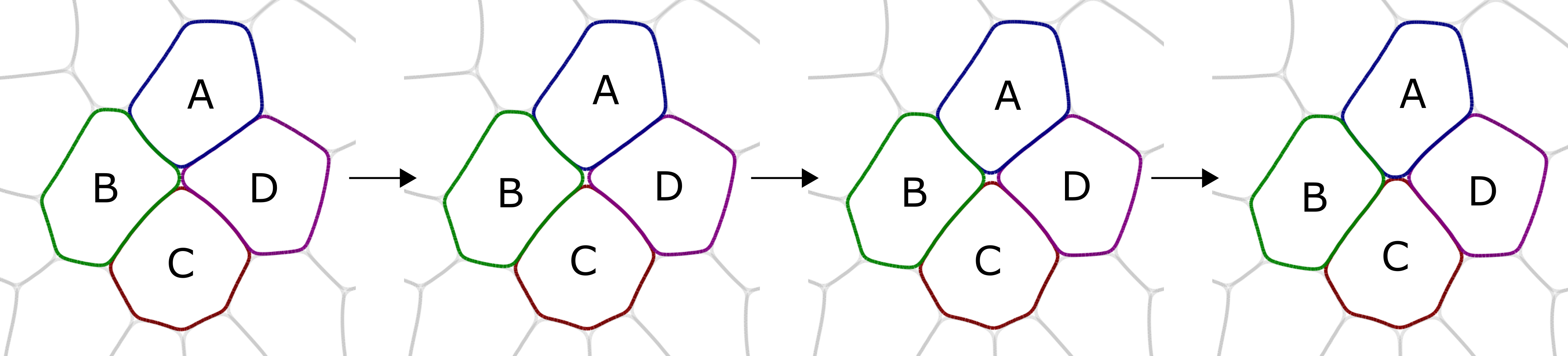}
    \caption{Successive time snapshots of tissue section undergoing a T1 transition, a finite-time neighbour exchange process between cells A, B, C and D. The transition starts when cells B and D lose contact and is completed when cells A and C make contact. During the T1 transition an extracellular space (gap) is formed between cells A, B, C and D. Also see Supplementary Movie 1}
    \label{fig: T1 transition}
\end{figure}

In confluent tissues, the tissue architecture can change in several ways. To isolate the tissue dynamics driven by spontaneous T1 transitions, we consider an idealised situation where apoptosis and cell division are neglected, cells have a constant volume, identical mechanical properties, and their total number is fixed. 
During a T1 transition, typically 
two neighbouring cells move apart, while two of their neighbours come towards each other and make contact as illustrated in Figure \ref{fig: T1 transition}. The average number of neighbours before and after the T1 transition is invariant. Through T1 transitions, cells undergo large deformations and shape changes, and encounter an energy barrier that they have to overcome through their activity~\cite{biDensityindependentRigidityTransition2015, oswaldJammingTransitionsCancer2017}. Albeit, there are several competing scenarios of the mechanical-chemical-biological feedback involved in a T1 transition, our understanding of these coupled processes is still elusive~\cite{rauziNatureAnisotropyCortical2008}. 

T1 transitions are common features also in granular matter and foams under external forcing \cite{schallStructuralRearrangementsThat2007, weairePhysicsFoams2001, stavansEvolutionCellularStructures1993}. The energy relaxation after a T1 transition has been studied in foams by measuring the length of T1 junctions \cite{durandRelaxationTimeTopological2006}. This concept was adapted for active tissues ~\cite{curranMyosinIIControls2017}, where the length of the T1 junction before and after a T1 transition has been measured. During a T1 transition in dry foams \cite{weairePhysicsFoams2001}, the cells form a rosette where either four or more edges meet. It has been shown that a junction is energetically stable for three edges incident at $120$ degrees. So, while undergoing a T1 transition, the cells pass from one metastable state to another via an unstable state comprising of a rosette. In confluent tissue extracellular spaces (gaps) change this process \cite{kimEmbryonicTissuesActive2021}. Rosettes and tri-junctions can no longer be defined by the number of edges meeting but are placed where gaps are formed. 

Also, various mathematical models have been used to study different facets of T1 transitions in foams and tissues \cite{biDensityindependentRigidityTransition2015,bartonActiveVertexModel2017,sknepnekGeneratingActiveT12022,kimEmbryonicTissuesActive2021, erdemci-tandoganEffectCellularRearrangement2021, drenckhanRheologyOrderedFoams2005}. They are mainly based on vertex models, which approximate cells by polygonal shapes. However, cell shape plays a crucial role in T1 transitions and the ability to accurately describe complex cell shapes, beyond polygons, might be advantageous
\cite{boromandJammingDeformablePolygons2018,perroneNonstraightCellEdges2016,kimEmbryonicTissuesActive2021}. We consider a multi-phase field model that allows for spontaneous T1 transitions while capturing the cell shape at a high resolution and allowing for large shape deformations. Multi-phase field models have been used to probe several questions pertaining collective motion of cells \cite{Nonomura2012,Palmieri2015,Marth2016,muellerEmergenceActiveNematic2019,loeweSolidLiquidTransitionDeformable2020,Camley2014,wenzelMultiphaseFieldModels2021,Wenzel2019,jainImpactContactInhibition2022}. These models consider cells as active incompressible droplets and unlike vertex models, T1 transitions emerge spontaneously as a result of shape deformations. In vertex models, also extracellular spaces (gaps) need explicit modelling with ad-hoc assumptions \cite{kimEmbryonicTissuesActive2021}, whereas in multi-phase field models they are emergent. 

In this paper, we focus on characterising the energy profile preceding and succeeding T1 transitions. We show that this energy profile is statistically robust to changes in several model parameters. It is characterized by an asymmetric profile with a fast increase in energy before the T1 transition, a sudden drop after the T1 transition, followed by a slow relaxation. The relaxation profile provides insights into the flow properties of the tissue. Previously, the relaxation has been indirectly studied in tissues by examining the relaxation of an ellipsoid droplet immersed in a tissue \cite{mongeraFluidtosolidJammingTransition2018,kimEmbryonicTissuesActive2021,durandRelaxationTimeTopological2006}. The relaxation profile was attributed to yield stress due to limitations in the measurement timescales \cite{mongeraFluidtosolidJammingTransition2018}, and also associated with the fluidization of the tissue \cite{kimEmbryonicTissuesActive2021,mongeraMechanicsCellularMicroenvironment2023}. We further consider the duration of T1 transitions and find that the average duration scales inversely to the maximum average energy attained during the T1 transition. Also, we show that T1 transitions may trigger the creation of other T1 transitions nearby and the chaining of T1 transitions leads to large-scale deformation and fluid like behaviour. 

We introduce the multi-phase field model in Section \ref{Sec:MultiPhaseField} and discuss results on local statistical properties of T1 transitions in Section \ref{Sec:Results}. We further analyse the dependency of these statistical properties on various model parameters. The effect of cell deformability and activity is considered in detail. We study the impact of chaining of T1 transitions on flow at larger scales. In Section \ref{Sec:Discussion} we relate these finding to mechanical and rheological properties of the tissue and postulate that they can be used to characterize fluidization. Details on the numerical methods, the initialization and characterization of T1 transitions are provided in Section \ref{Sec:Methods}. 
%%%%%%%%%%%%%%%

%-------------------
\section{Multi-phase field model}\label{Sec:MultiPhaseField}

We represent a two-dimensional confluent cell tissue within a multi-phase field model following formulations \cite{Marth2016, Wenzel2019, wenzelMultiphaseFieldModels2021, jainImpactContactInhibition2022}. We consider a system of $N$ cells of equal area occupying a square domain of size $[0,L] \times [0,L]$ and use periodic boundary conditions. Each cell is represented by a scalar phase field $\phi_i(\textbf{x}, t)$ as an indicator function of the domain occupied by each cell labeled by $i = 1, 2,\cdots, N$. Namely, the bulk phase values $\phi_i \approx 1$ and $\phi_i \approx -1$ indicate the interior and exterior of the cell, respectively. The cell boundary is defined by the localised transition region between the two bulk values. The time evolution of the $i$-th phase field follows a conservative dynamics which preserves the cell areas and is given by 
    \begin{equation}\label{eq: Phase field evolution}
        \partial_t \phi_i + \mathbf{v}_i \cdot \nabla \phi_i = \Delta \frac{\delta  \mathcal{F}}{\delta \phi_i},
    \end{equation}
 where $\Delta$ is the two-dimensional Laplacian applied to the variational derivative of a free energy functional $\mathcal{F}$ with respect to the phase field $\phi_i$. The free energy $\mathcal{F} = \mathcal{F}_{CH}  + \mathcal{F}_{INT} $ contains the Cahn-Hilliard energy 
    \begin{equation}\label{eq: Cahn Hilliard Energy}
        \mathcal{F}_{CH} = \frac{1}{Ca}\sum_{i=1}^N \int_{\Omega} \left(\frac{\epsilon}{2}||\nabla \phi_i||^2 + \frac{1}{4\epsilon}(\phi_i^2-1)^2\right) d\mathbf{x},
    \end{equation}
and the interaction energy \cite{Marth2016,jainImpactContactInhibition2022}
    \begin{equation}\label{eq: Interaction Energy}
        \mathcal{F}_{INT} = \frac{1}{In}\sum_{i=1}^N \int_{\Omega} B(\phi_i)\sum_{j\neq i} w(\phi_j) d\mathbf{x}.
    \end{equation}
The capillary number $Ca$ and interaction number $In$ are tuning parameters for the cell deformability and the strength of mutual repulsion/attraction interactions, respectively. 
In equation~\ref{eq: Cahn Hilliard Energy}, the Cahn-Hilliard energy has a local free energy density given by the double well potential with the minima corresponding to the two bulk values and a gradient energy. The parameter $\epsilon$ controls the width of the diffuse interface. The Cahn-Hilliard energy ensures phase separation into two bulk regions which are separated by a thin, diffusive interface. This energy alone is minimised by cells with circular shapes. In equation~\ref{eq: Interaction Energy}, each cell's interior and interface ($B(\phi_i) = (\phi_i+1)/2$) is coupled with every other cell through a local interaction potential,
    \[w(\phi_j) = 1 - (a+1) \left(\frac{\phi_j - 1}{2}\right)^2 + a\left(\frac{\phi_j - 1}{2}\right)^4,\] 
 where the parameter $a=1$ models repulsion, while $a>1$ models attraction and repulsion (see \cite{jainImpactContactInhibition2022} for a detailed analysis of role of $a$). 
    
Cell activity is introduced through the advection velocity $\mathbf{v}_i(\mathbf{x}, t)$ in equation \ref{eq: Phase field evolution} and is given by  
    \begin{equation}
        \mathbf{v}_i(\mathbf{x}, t) = v_0 B(\phi_i) \mathbf{e}_i(t),
    \end{equation} 
    where $v_0$ is a constant parameter that controls the magnitude of the activity, $\mathbf{e}_i= [\cos \theta_i (t), \sin  \theta_i(t) ]$ where $\theta_i$ is the orientation of the self-propulsion which evolves as 
    \begin{equation}\label{eq: evolution of theta}
        d\theta_i = \sqrt{2 D_r }dW_i(t) + \alpha(\beta_i(t) - \theta_i(t))dt.
    \end{equation}
    The first term on the right side of equation~\ref{eq: evolution of theta} is a rotational diffusion term with a Wiener process $W_i$. The second term is a relaxation to the orientation of the cell's shape elongation. The cell elongation is identified by the principal eigenvector of the shape deformation tensor \cite{muellerEmergenceActiveNematic2019, wenzelMultiphaseFieldModels2021} 
    \begin{equation}
        \mathbf{S}_i = \begin{bmatrix} S_{i,0} & S_{i,1} \\ S_{i,1} & -S_{i,0}\end{bmatrix}
    \end{equation}
which is symmetric and traceless and has the two components 
    \begin{equation*}
        S_{i,0} = \frac{1}{8}\int_\Omega \left[\left(\frac{\partial \phi_i}{\partial y}\right)^2-\left(\frac{\partial \phi_i}{\partial x}\right)^2\right]d\mathbf{x} \quad \mbox{and} \quad 
        S_{i,1} = -\frac{1}{4}\int_\Omega \left[\frac{\partial \phi_i}{\partial x}\frac{\partial \phi_i}{\partial y}\right]d\mathbf{x}.
    \end{equation*}
     Its corresponding eigenvalues are $\lambda^{\pm}_i = \pm \sqrt{S_{i,0}^2+S_{i,1}^2}$ and eigenvectors are $\mathbf{\eta}_i^{\pm} = ( \frac{S_{i,0} + \lambda_i^{\pm}}{S_{i,1}} ,  1)$. The vector $\mathbf{\eta}_i^{+}$ is parallel to the elongation axis of the cell and determines the preferred self-propulsion direction as  
     \begin{equation}
     \beta_i(t) = \left\{ 
     \begin{array}{rl}
     \arg(\boldsymbol{\eta}_i^{+}(t)) &: \;\; \mathbf{e}_i(t)\cdot \mathbf{\eta}_i^{+}(t) > 0 \\ -\arg(\boldsymbol{\eta}_i^{+}(t)) &: \;\; \mathbf{e}_i(t)\cdot \mathbf{\eta}_i^{+}(t) < 0 
     \end{array}\right.
     \end{equation}
     Therefore, the second term on the right hand side of equation \eqref{eq: evolution of theta} aligns $\theta_i(t)$ with $\beta_i(t)$. The parameter $\alpha$ controls the time scale of this alignment of the self-propulsion direction with the elongation axis of the cell. There are different possibilities to define the advection velocity $\mathbf{v}_i(\mathbf{x}, t)$ (see Ref. \cite{wenzelMultiphaseFieldModels2021} for an overview and comparison). The current form includes approaches of Ref. \cite{muellerEmergenceActiveNematic2019} 
     and, as the elongation is a result of the interaction with neighbouring cells, it accounts for contact inhibition of locomotion \cite{smeetsEmergentStructuresDynamics2016,stramerMechanismsVivoFunctions2017}. The model leads to properties appropriate to describe, e.g., Madin-Darby canine kidney
(MDCK) cells \cite{peyretSustainedOscillationsEpithelial2019,wenzelMultiphaseFieldModels2021}.

%%%%%%%%%%% Figure %%%%%%%%%%%%%
%%%%%%%%%%%%%%%%%%%

%%%%%%%%%% Figure -----------
\begin{figure}[htb!]
    \centering
    \begin{subfigure}[b]{0.3\textwidth}
        \centering
        \includegraphics[width=\textwidth]{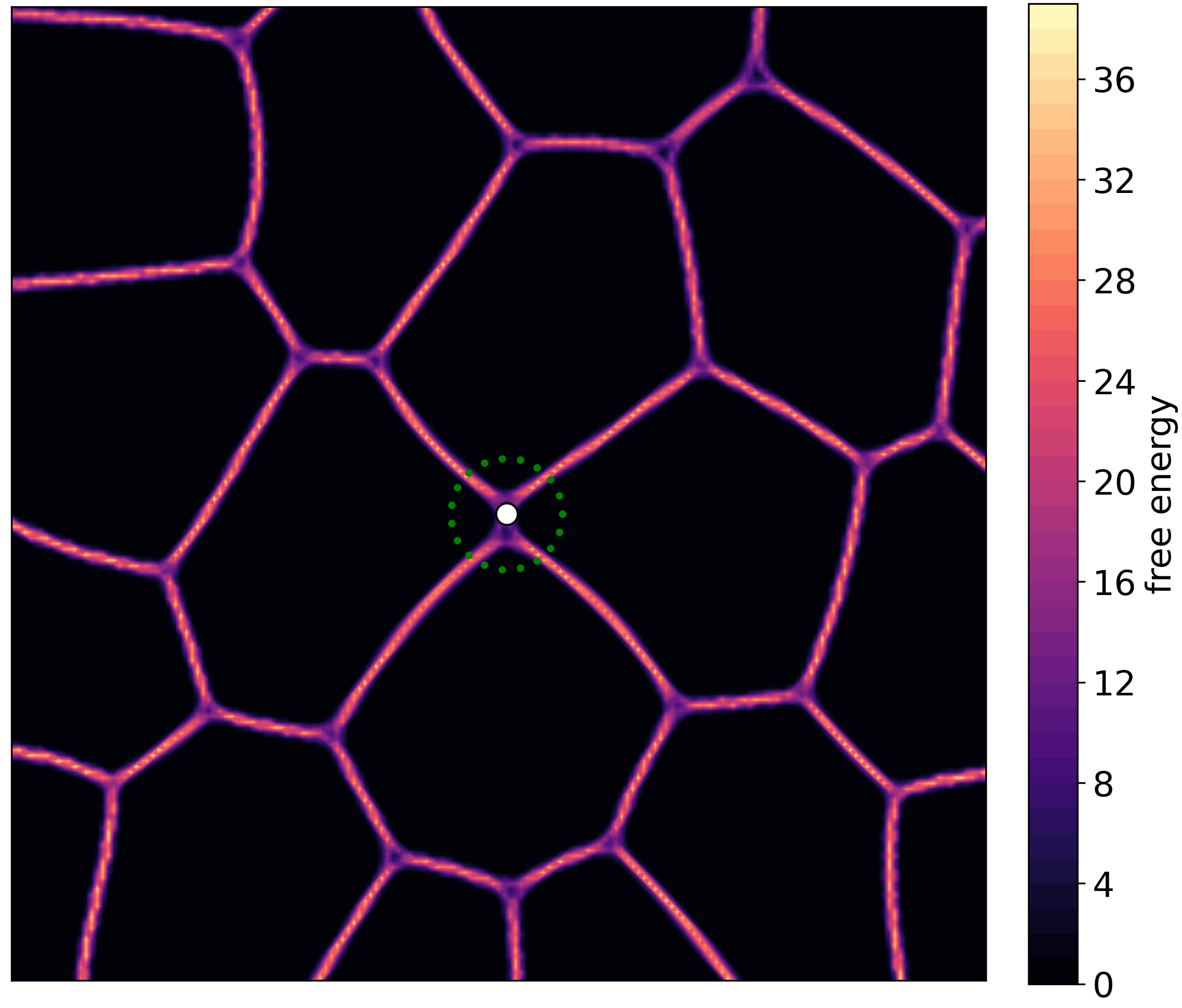}
        \caption{}
        \label{fig: free energy}
    \end{subfigure}
    %\hfill
    \begin{subfigure}[b]{0.3\textwidth}
        \centering
        \includegraphics[width=\textwidth]{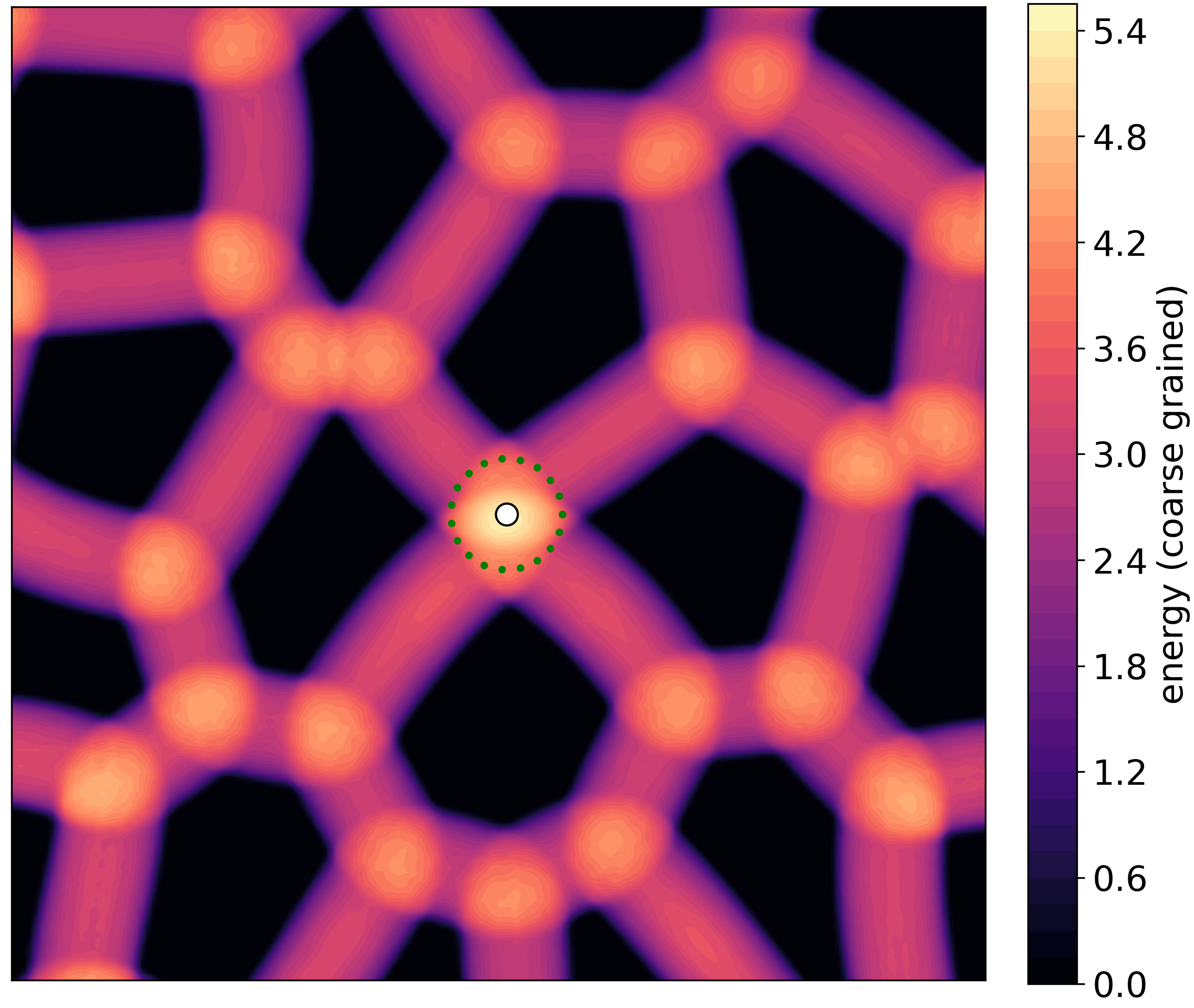}
        \caption{}
        \label{fig: averaged free energy}
    \end{subfigure}
    %\hfill
    \begin{subfigure}[b]{0.3\textwidth}
        \centering
        \includegraphics[width=\textwidth]{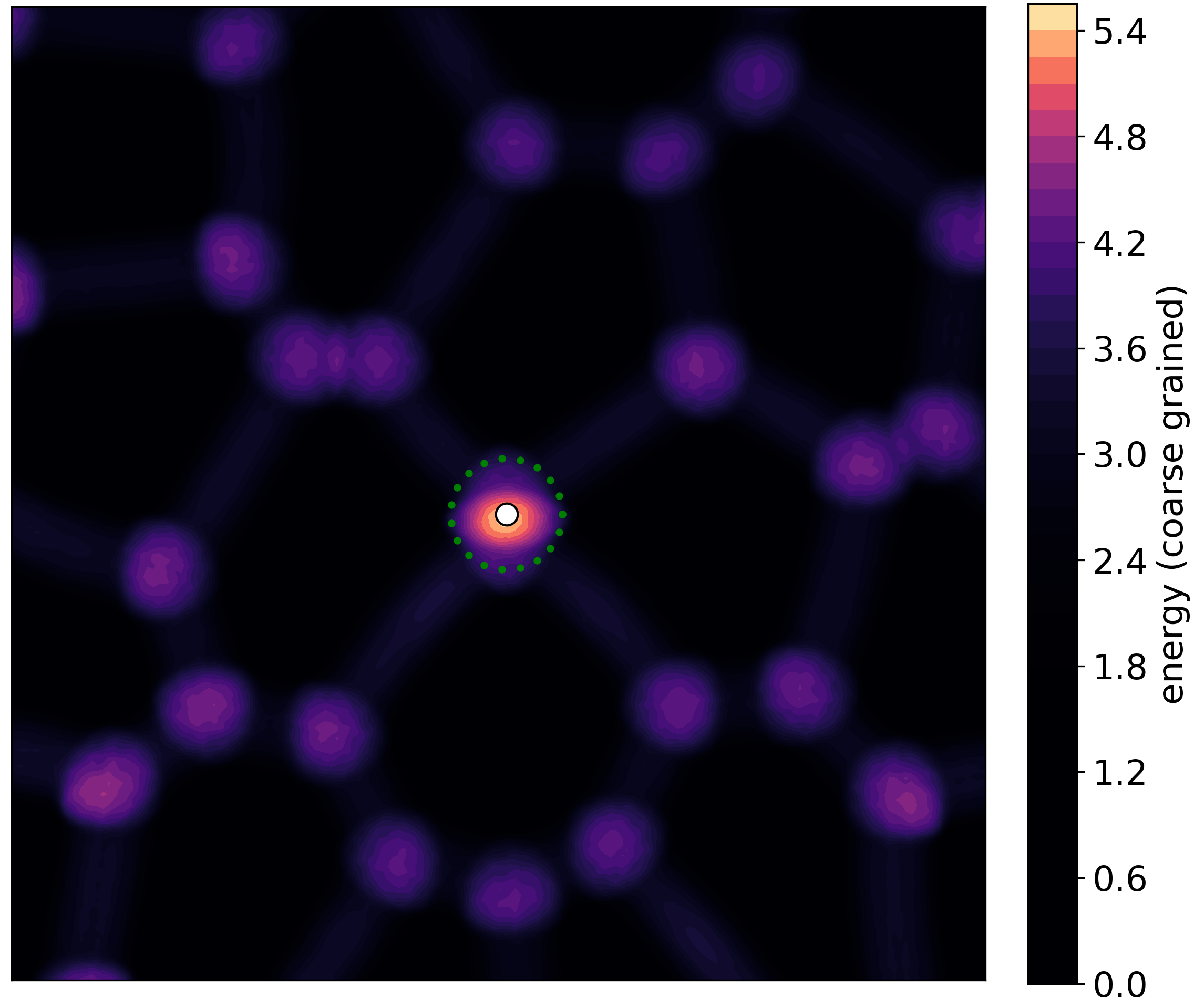}
        \caption{}
        \label{fig: averaged free energy in power law scale}
    \end{subfigure}
    \caption{(a). Free energy density, in a region surrounding a T1 transition. (b) and (c) Coarse-grained energy density in a linear and log-scale, respectively. The white dot represents the epicenter of the T1 transition while the green dotted circle represents the coarse graining radius $r_{avg}$, the estimated core of the T1 transition.}
    \label{fig: Energy averaging}
\end{figure}
%%%%%%%%%%&&&&&&&&&&&

%%%%%%%%%%%%%%%%%%%%%%%
\section{Results}\label{Sec:Results}

\subsection{Energy profile of T1 transitions}

Within our multi-phase field approach, T1 transitions are neighbour exchange processes with a finite duration. A prototypical time sequence of a T1 transition is illustrated in Figure~\ref{fig: T1 transition}. Four cells A, B, C and D are involved. Before the T1 transition, the cell junction shared by cells B and D shrink. The T1 transition starts when the cells B and D break contact and move apart. This results in the formation of an extracellular space which we call 'gap'. Cells A and C move towards each other, close the gap, and form a new contact concluding the T1 transition. After the T1 transition, this new junction between cells A and D  expands. The junctions that shrink and expand are called T1 junctions. We refer to Section \ref{Sec:Methods} for the procedure to detect T1 transitions and their durations. A T1 transition not only leads to topological rearrangements of the four neighbouring cells, it also involves deformation of the cells. While details, such as the specific shape of the cells and their deformation, the duration of the T1 transition and the relaxation process differ between T1 transitions, we will demonstrate that robust statistical features of T1 transitions exist. 

%%%%%%%%%% Figure &&&&&&&&&&&&&&&&
\begin{figure*}[!ht]
    \includegraphics[width=0.97\textwidth]{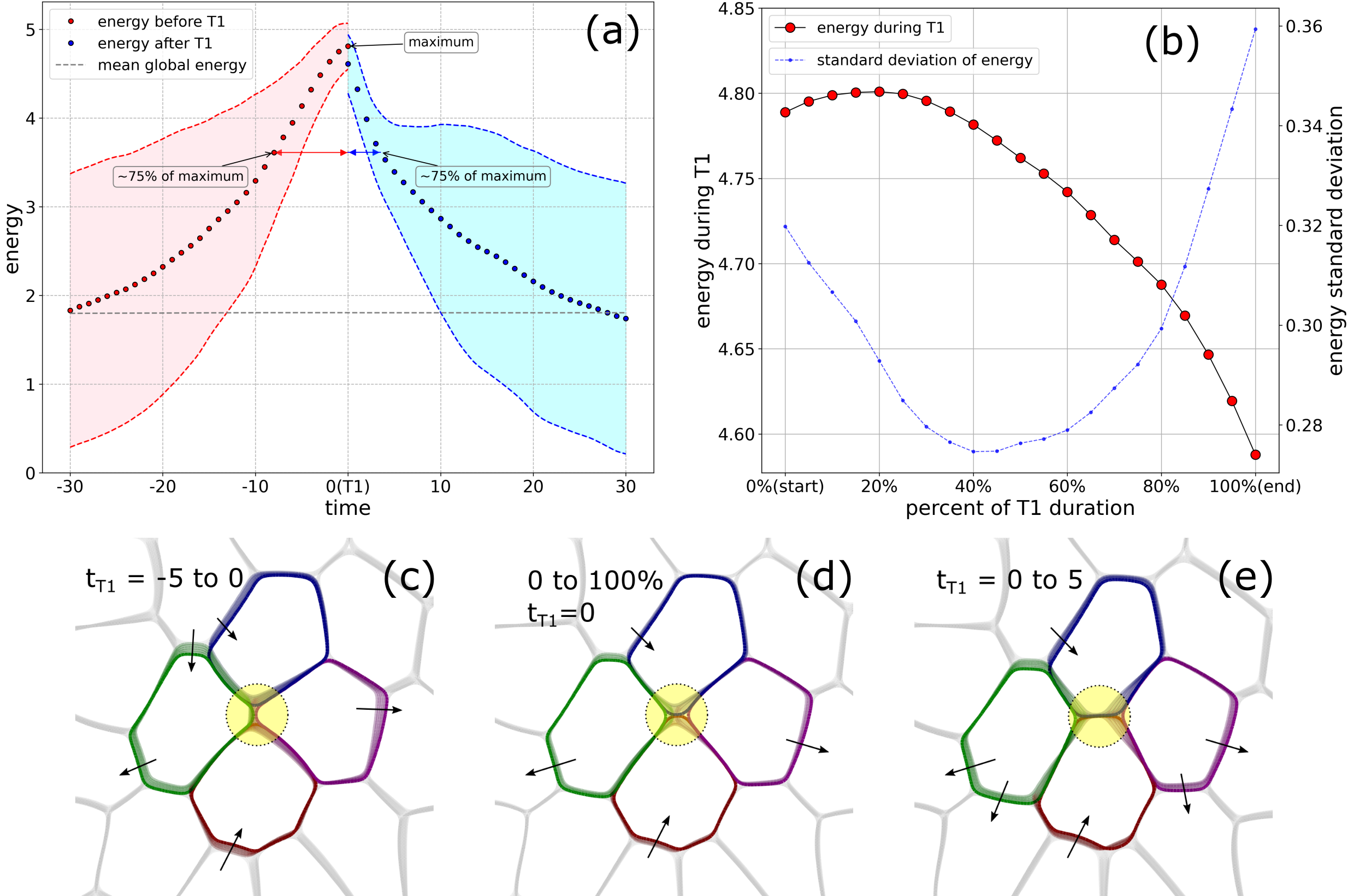}
    \caption{(a) Evolution of energy (averaged for 158 T1 transitions) at the epicenter of the T1 transitions. Negative time corresponds to time before a T1 transition and positive time corresponds to time after a T1 transition. The shaded region denotes a width of 1 standard deviation. The gray dashed line is the average energy across the whole domain. (b) Average energy profile during a T1 transition as function of percentage of T1 duration. The standard deviation is also indicated. (c), (d) and (e) Montages of deformed cells involved in a T1 transition. Each montage is made up of 5 images, that capture the cells at equidistant times, stacked over each other. The darkest colored overlay represents the latest time. (c)  Cell shapes before the start of the T1 transition, (d) during the T1 transition, and (e) after the end of the T1 transition. Also see Supplementary Movie 2 for corresponding simulation}
    \label{fig: energy prepost profile}
    \label{fig: energy during T1}
    \label{fig: mosaic before}
    \label{fig: mosaic during}
    \label{fig: mosaic after}
\end{figure*}
%%%%%%%%%%&&&&&&&&&&&
We define the epicenter of a T1 transition as the point with the minimum total distance from the centers of the involved cells in the neighbour exchange process midway through the T1 transition. We define the immediate region around the epicenter as the core of a T1 transition, which is of essence because it is the region where T1 junctions shrink and expand, and the gap appears and disappears. 

Figure~\ref{fig: free energy} shows the total free energy density midway through a T1 transition. The epicentre is shown by the white dot and the estimated core is highlighted by the green circle. It has a radius $r_{avg} = 0.02L$, where $L$ is the side length of the computational domain. We compute a coarse-grained energy whose value at any point in the domain is the average of the energy density in a circular region centered at that point with radius $r_{avg}$. Figure~\ref{fig: averaged free energy} shows this coarse grained energy field $f_{r_{avg}}$, which we will call 'energy' field in the following. The signature of triple-junctions and T1 transitions already becomes appealing due to their higher energy. The difference between both is enhanced by using a log scale, see Figure~\ref{fig: averaged free energy in power law scale}. Considering this energy field in the epicenter over time provides a spatial-temporal description of T1 transitions. For discussions on the sensitivity of this procedure on $r_{avg}$ we refer to Section \ref{Sec: Sensitivity to coarse graining}.

Figure~\hyperref[fig: energy prepost profile]{3a} shows the time evolution of this energy averaged over 158 T1 transitions. The time is negative before a T1 transition and is positive after a T1 transition, and is denoted by $t_{T1}$. The energy during the T1 transitions is excluded, which leads to a discontinuity at $t_{T1} = 0$. The two values at $t_{T1}=0$ correspond to the averaged energies at the start and the end of the T1 transitions. As the duration of T1 transitions differs, an averaged energy as a function of time during the T1 transition does not provide any meaningful information. Details on the energy during the T1 transition are shown in Figure~\hyperref[fig: energy during T1]{3b} using a normalized time. The energy profile in 
Figure~\hyperref[fig: energy prepost profile]{3a, 3b} has a peak at the T1 transition. The profile is asymmetric with a strong increase in energy before the T1 transition and a sudden decrease after the T1 transition followed by a slow relaxation. The asymmetry can be quantified by considering the $75\%$ of the maximum value, which is marked in Figure \hyperref[fig: energy prepost profile]{3a}. Figures~\hyperref[fig: mosaic before]{3c-3e} illustrate the evolution for one T1 transition, the one depicted in Figure~\ref{fig: T1 transition}. These figures contain overlays of several snapshots as per the time marked in the figures. The darkest of these snapshots pertains to the latest time. The yellow region marks the estimated core of the T1 transition. The asymmetry before and after the T1 transition, Figure~\hyperref[fig: mosaic before]{3c, 3e}, respectively, is clearly visible. The T1 junctions are longer at $t_{T1} = 5$ compared with $t_{T1} = -5$. During the T1 transition, Figure~\hyperref[fig: mosaic before]{3d}, the asymmetry is less pronounced. Most of the deformations are concentrated in the core. These deformations arise as a result of the formation of the gap, and subsequently its disappearance. The shrinking and formation of T1 junctions and the deformations within the core are a signature of the T1 transition. However, they also influence the deformation of the four cells outside of the core, and their neighbours, which can be perceived by the overlayed cell shapes. Interestingly in the depicted T1 transition, the deformations of each of the four cells seems to be persistent before, during and after the T1 transition (see the arrows indicating the direction of deformations). We will elaborate on this and other coarse grained effects in Section \ref{Sec: T1 Chain}. 
The energy profile indicates an accumulation of energy to reach the energy barrier at the T1 transition. This is due to probing several possibilities in local movement and cell shape deformation, which are coupled by the definition of activity, taking into account cell elongation and contact inhibition of locomotion. After the energy barrier has been overcome the fast relaxation of the energy can be associated with a steep gradient in the energy landscape in one direction. 
%The energy profile indicates an accumulation of energy to reach the energy barrier at the T1 transition. Deformations arise in the tissue as a result of the shape dependent activity of the cells whose shape is constrained by the shape of their neighbours. Due to the feedback between cell activity and shape, a variety of possible cell shapes are explored, some of which leads to the energy accumulation at the core region. Eventually, the energy barrier is overcome through a neighbour exchange resulting in the fast relaxation of the energy, which is associated with a steep gradient in the energy landscape in one direction. Figure~\hyperref[fig: mosaic before]{3e} shows that the portion within the core has significant deformations compared to outside the core. These large but localised deformations are attributed to fast energy relaxation rather than activity. 
 
%%%%%%%%%% Figure &&&&&&&&&&&
\begin{figure}[!ht]
    \centering    
    \includegraphics[width=0.85\linewidth]{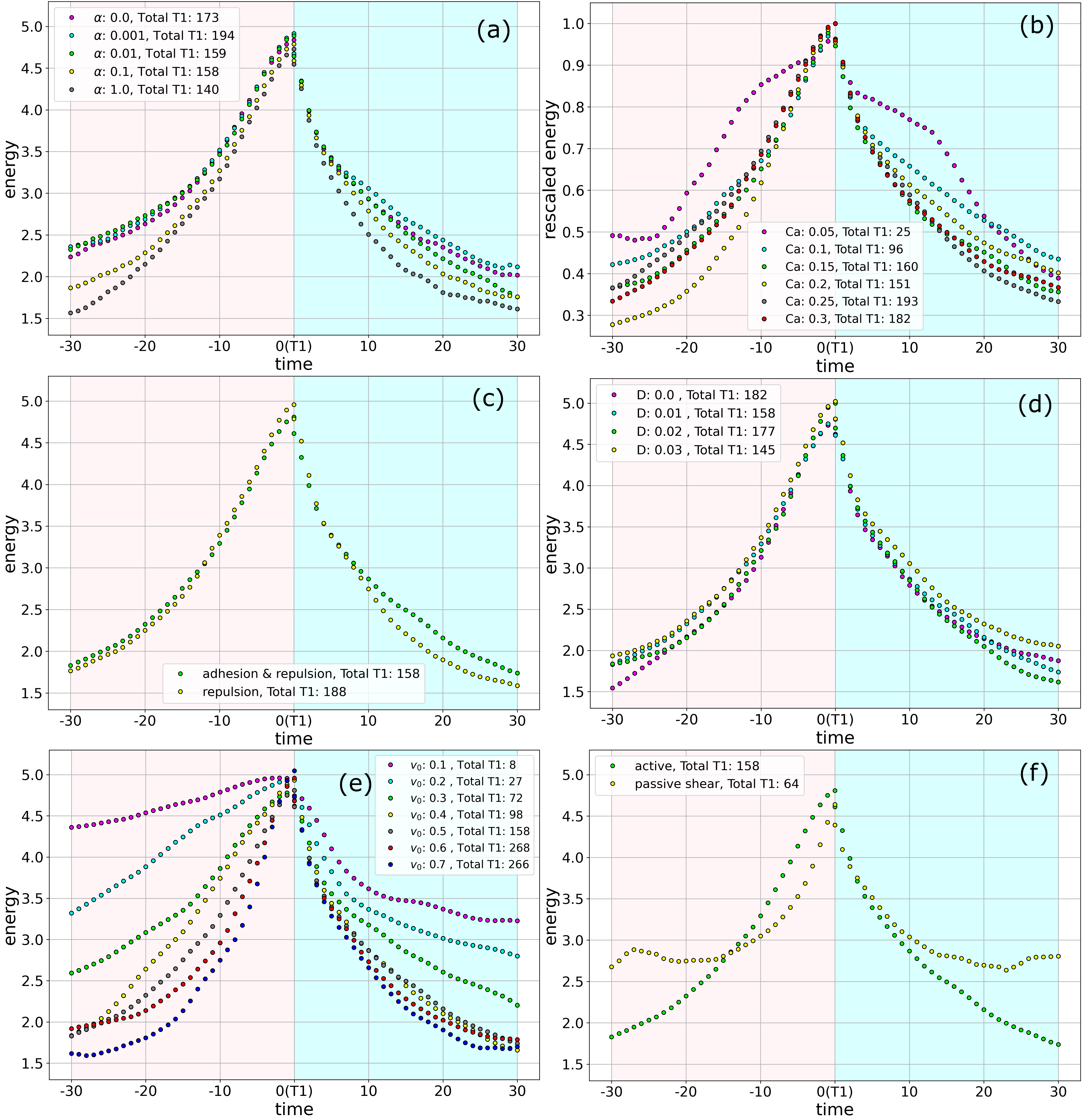}
    \caption{Evolution of energy for different parameter values. The pink and cyan shaded region are used to denote time before and after the T1 transitions, respectively. The number of T1 transition used to obtain these results in indicated. (a) The aligning parameter $\alpha$ is varied. (b) The parameter to control cell deformability, $Ca$ is varied. As $Ca$ is a parameter that influences the overall total energy, for better comparison the energy is rescaled by division with the maximum energy. (c) Adhesion and repulsion corresponds to $a=1.5$ and repulsion corresponds to $a=1$. (d) The diffusivity $D$ is varied. (e) The magnitude of the activity $v_0$ is varied. (f) The passive shear corresponds to advection field $v_i(\mathbf{x},t) = 0.5  |x_1-\frac{L}{2}|$ while the active case corresponds to parameters in Table~\ref{tab: parameters}.}
    \label{fig: T1 transition energy profiles}
    \label{fig: energy prepost Ca}
    \label{fig: energy prepost interaction}
    \label{fig: energy prepost alpha}
    \label{fig: energy prepost diffusion}
    \label{fig: energy prepost velocity}
    \label{fig: energy prepost shear}
\end{figure}
%%%%%%%%%%&&&&&&&&&&&

The asymmetric shape of the energy profile is robust to changes in most model parameters, as demonstrated in Figure~\ref{fig: T1 transition energy profiles} where $\alpha$, $Ca$, $a$, $D$ and $v_0$ are varied and the energy profile associated with passive sheared foams is included for comparison. Figure \hyperref[fig: energy prepost Ca]{4b} shows the energy rescaled by the maximum energy as changes in $Ca$ directly affect the free energy, see equation \eqref{eq: Cahn Hilliard Energy}. Within the range of parameters explored, the changes in the values of alignment parameter $\alpha$, interaction coefficient $a$, and diffusivity $D$ have minimal effects on the energy profile. We see that the profile is robust even in absence of noise ($D=0$) (Figure~\hyperref[fig: energy prepost diffusion]{4d}). On the other hand, the profile deviates from Figure \hyperref[fig: energy prepost profile]{3a} for low values of $v_0$ and $Ca$. Figure~\hyperref[fig: energy prepost velocity]{4e} shows that the cell activity $v_0$ affects the rate at which the cells approach a T1 transition which is indicated by the slower accumulation of energy for low $v_0$. However, change in $v_0$ has a minor effect on the energy relaxation immediately after a T1 transition. The slow relaxation afterwards is largest for large values of $v_0$. This can be associated with the definition of activity, which is related to cell elongation and at least on average cells elongate in the direction of movement after the T1 transition. The characteristic profile of the accumulation of energy before the T1 transition and the fast relaxation of energy after the T1 transition is also present for low values of $Ca$, see Figure~\hyperref[fig: energy prepost velocity]{4b}. However, as Figure~\hyperref[fig: energy prepost velocity]{4b} considers a rescaled energy the actual rates depend on $Ca$. The slow relaxation after the sudden decrease only slightly depends on $Ca$. We would like to point out that the results for low values of $v_0$ and $Ca$ should be considered with care, as the number of T1 transitions considered in these cases is much lower. While the system is still in the fluid phase, the extreme values for $v_0 = 0.1$ and $Ca = 0.05$ already approach the transition to the solid phase. 

In passive foams T1 transitions can be induced by applying shear. This is considered by an advection velocity field $v_i(\mathbf{x},t) = 0.5  |x_1-L/2|$ and the resulting energy profile is compared with the profile from Figure \hyperref[fig: energy prepost profile]{3a}, see Figure~\hyperref[fig: energy prepost shear]{4f}. 
The profiles differ before the T1 transition and within the slow relaxation, but are similar in the sudden drop of energy right after the T1 transition. The latter reiterates that the energy relaxation right after a T1 transition is independent on activity. 
%This is further exemplified by the Figure~\hyperref[fig: mosaic after]{3e} where bulk of the cells move relatively less while the portion of the cells within the core deforms significantly. This significant deformation is primarily controlled by cell deformability rather than by activity. This also results in higher velocity of the center of mass of the cells which will be discussed in section \ref{sec: duration and other properties}. 
The differences in the accumulation of the energy can be associated with the persistent orientation of advection velocity due to shear, which results in collective deformation and a more deterministic approach of the T1 transition. Also the termination of the decay in the passive case results from the restricted possibilities of relaxation due to the applied shear. 

%%%%%%%%%% Figure &&&&&&&&&&&
\begin{figure}[htb!]
    \centering
    \begin{subfigure}[b]{0.35\textwidth}
        \centering
        \includegraphics[width=\textwidth]{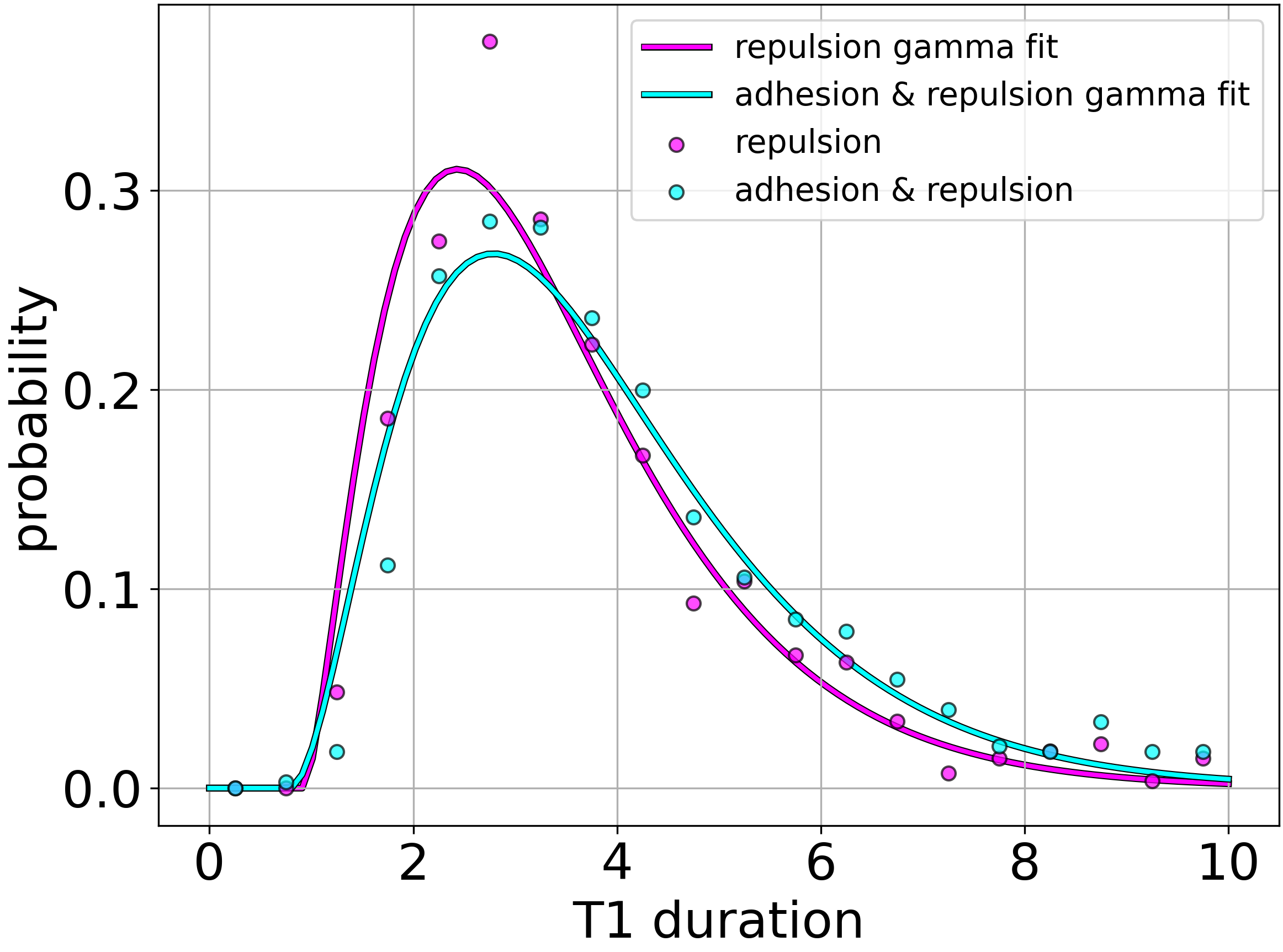}
        \caption{}
        \label{fig: T1 duration}
    \end{subfigure}
    %\hfill
    \begin{subfigure}[b]{0.35\textwidth}
        \centering
        \includegraphics[width=\textwidth]{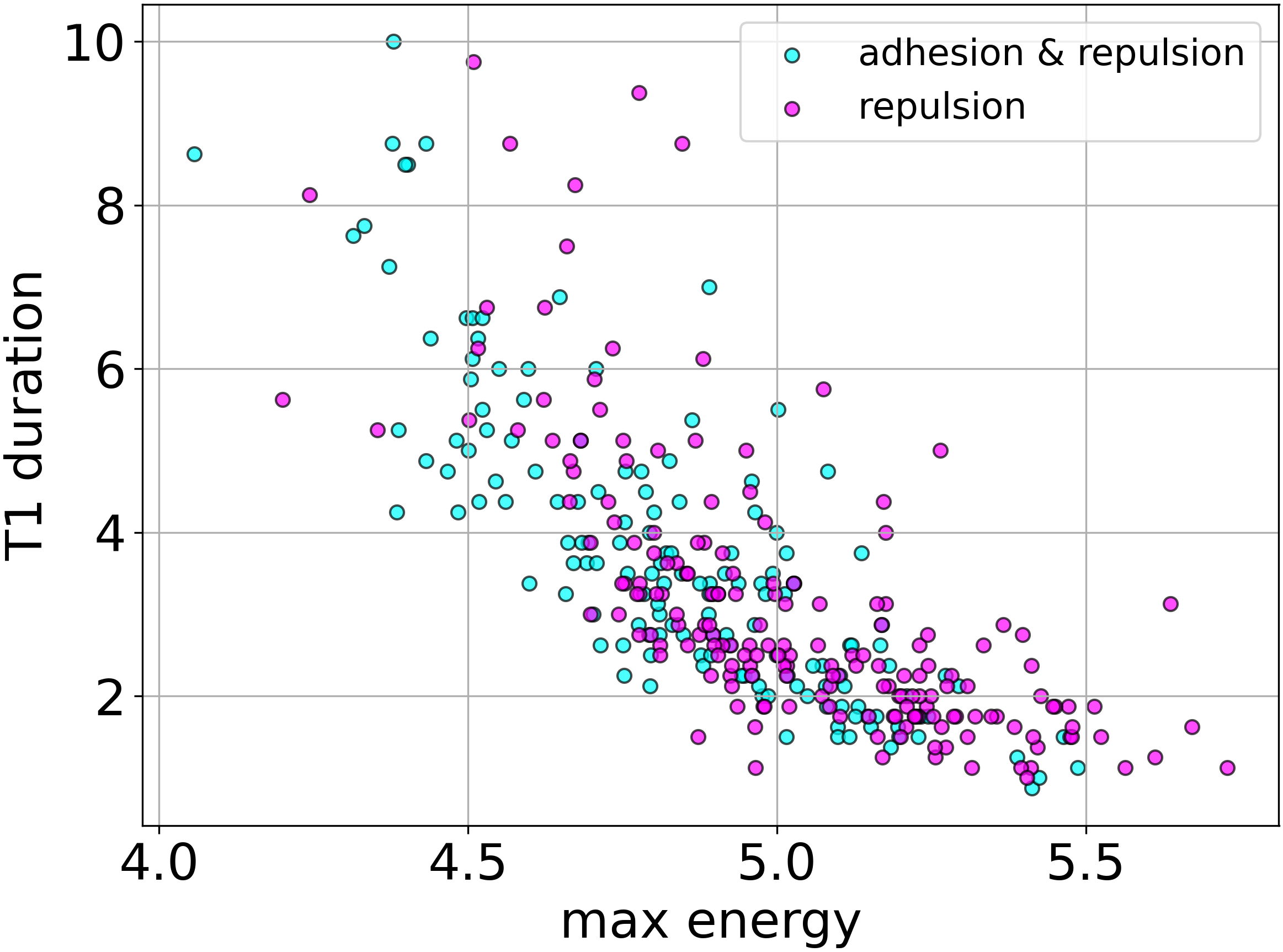}
        \caption{}
        \label{fig: T1 duration and energy}
    \end{subfigure}
    %\hfill
    \begin{subfigure}[b]{0.35\textwidth}
        \centering
        \includegraphics[width=\textwidth]{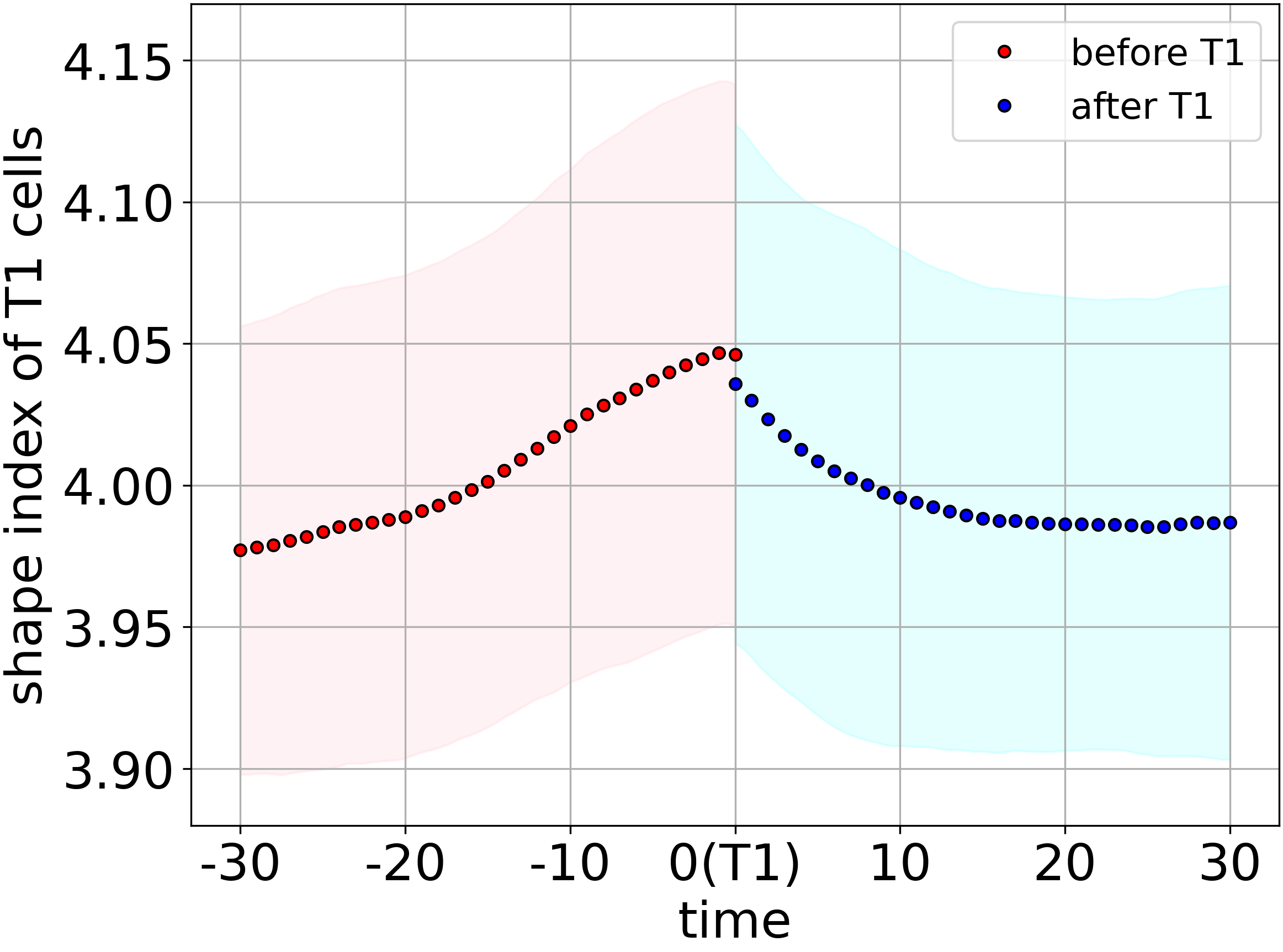}
        \caption{}
        \label{fig: shape index T1 }
    \end{subfigure}
    \begin{subfigure}[b]{0.35\textwidth}
        \centering
        \includegraphics[width=\textwidth]{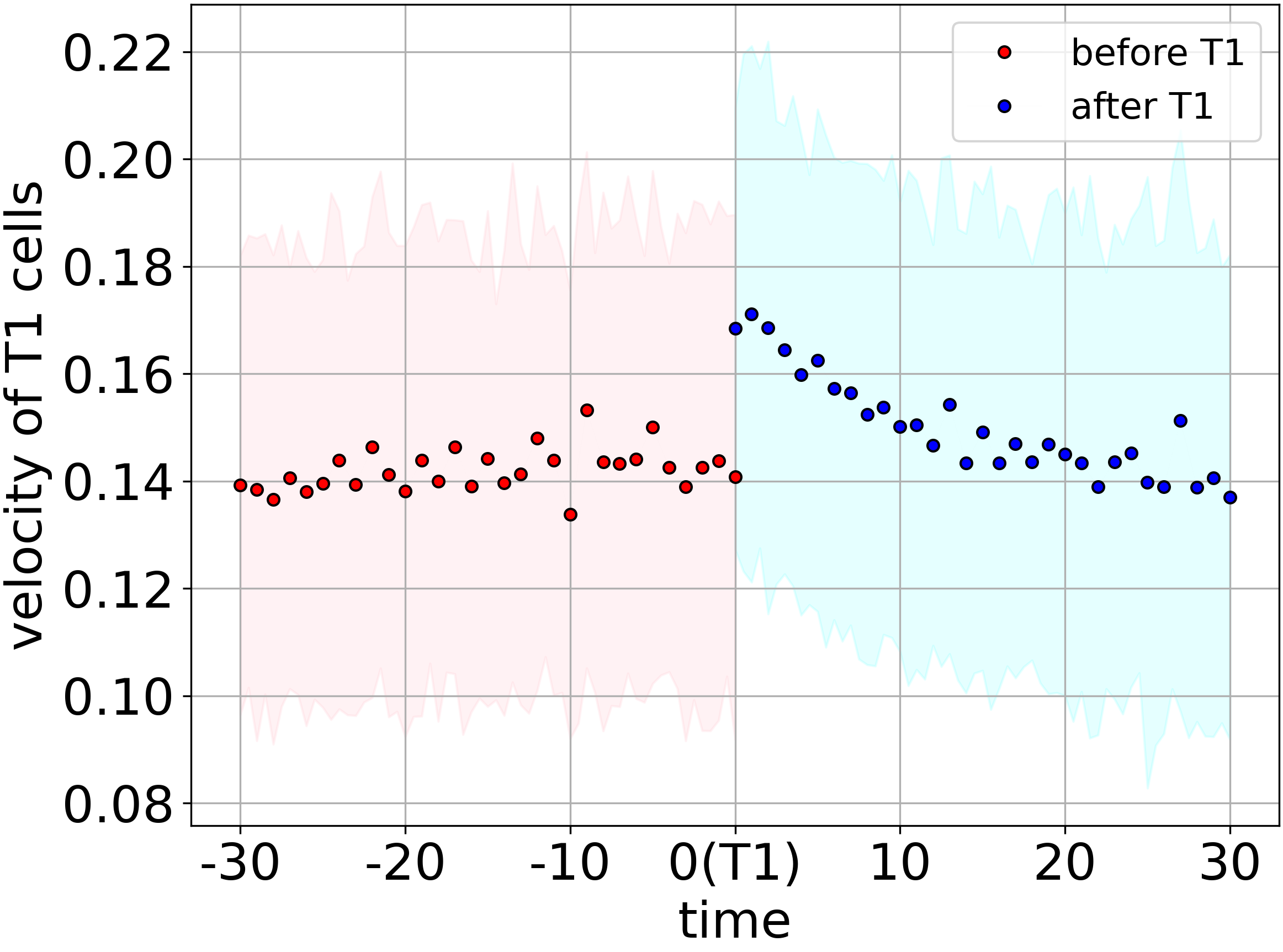}
        \caption{}
        \label{fig: velocity T1}
    \end{subfigure}
    \caption{(a) Probability distributions of the duration of T1 transitions for only repulsion interactions (magenta dots) and for both repulsion and adhesion (cyan dots). Both data sets are fitted by Gamma distributions highlighting the exponential tails. (b) Scatter plot of duration of T1 transition as function of the maximum energy reached during a T1 transition. (c) Evolution of average shape index and (d) Evolution of the average velocity of center of mass of the cells involved in the T1 transitions as function of time relative to a T1 transition. The shaded regions mark the standard deviations of both quantities.}
    \label{fig: T1 transition shape index, velocity, duration and max energy vs duration}
\end{figure}
%%%%%%%%%%&&&&&&&&&&&

\subsection{Duration and other properties of T1 transitions}\label{sec: duration and other properties}
As mentioned earlier, the duration of T1 transitions strongly depends on the specific cell arrangements. We now discuss the statistical properties of the duration. Figure~\ref{fig: T1 duration} shows the probability distributions of the duration of T1 transitions. The distributions peak at smaller values and have a long tail for larger values. The profiles corresponds to repulsive and adhesive ($a>1$),  and only repulsive interactions ($a=1$), and are fitted by Gamma distributions. The average duration of T1 transitions for repulsive interactions (3.418 measured for 539 T1 transitions across 3 simulations) is smaller compared to that for repulsive and adhesive interactions (3.826 for 631 T1 transitions across 4 simulations). Keeping other parameters fixed, the average number of T1 transitions in the repulsive and adhesive case was 157.75 while for the repulsive case was 179.66, respectively. Therefore, in the repulsive case, cells undergo neighbour exchanges faster and more often. Figure \ref{fig: T1 duration and energy} shows the  duration of T1 transitions as a function of the maximum energy reached during a T1 transition. While the data is scattered, it qualitatively shows that high energy T1 transitions are faster. This qualitative result holds for both cases and can be explained by a larger accumulation of energy in the core, which increases the spatial energy gradients and in turn speeds up the relaxation of the energy which leads to the shorter duration.

Figure \ref{fig: shape index T1 } shows the averaged shape index (perimeter$/\sqrt{\text{area}}$) of the four cells involved in a T1 transition as function of time relative to a T1 transition. The asymmetry found for the energy profile and the discontinuity at $t_{T1} = 0$ is also present for this quantity. The cells deform and elongate as they approach a T1 transition and relax afterwards. This increases and decreases their shape index, respectively. The faster relaxation leads to the asymmetry in the evolution of the shape index. The asymmetry around a T1 transition is also seen in the average velocity of the center of mass of the  cells involved in a T1 transition as shown in Figure \ref{fig: velocity T1}. While the velocity is almost constant before the T1 transition, the velocity peaks at the T1 transition and slows down afterwards until it reaches the average value before the T1 transition. The peak in the average velocity of the center of mass is due to the large deformations of the portions of cells within the core and their fast relaxation after the T1 transition. Both quantities, the shape index and the cell velocity of the four cells involved in a T1 transition are also experimentally accessible. These quantities can be related to the energy considered above.

%%%%%%%%%% Figure &&&&&&&&&&&
\begin{figure}[!htb]
    \centering
    \includegraphics[width=\textwidth]{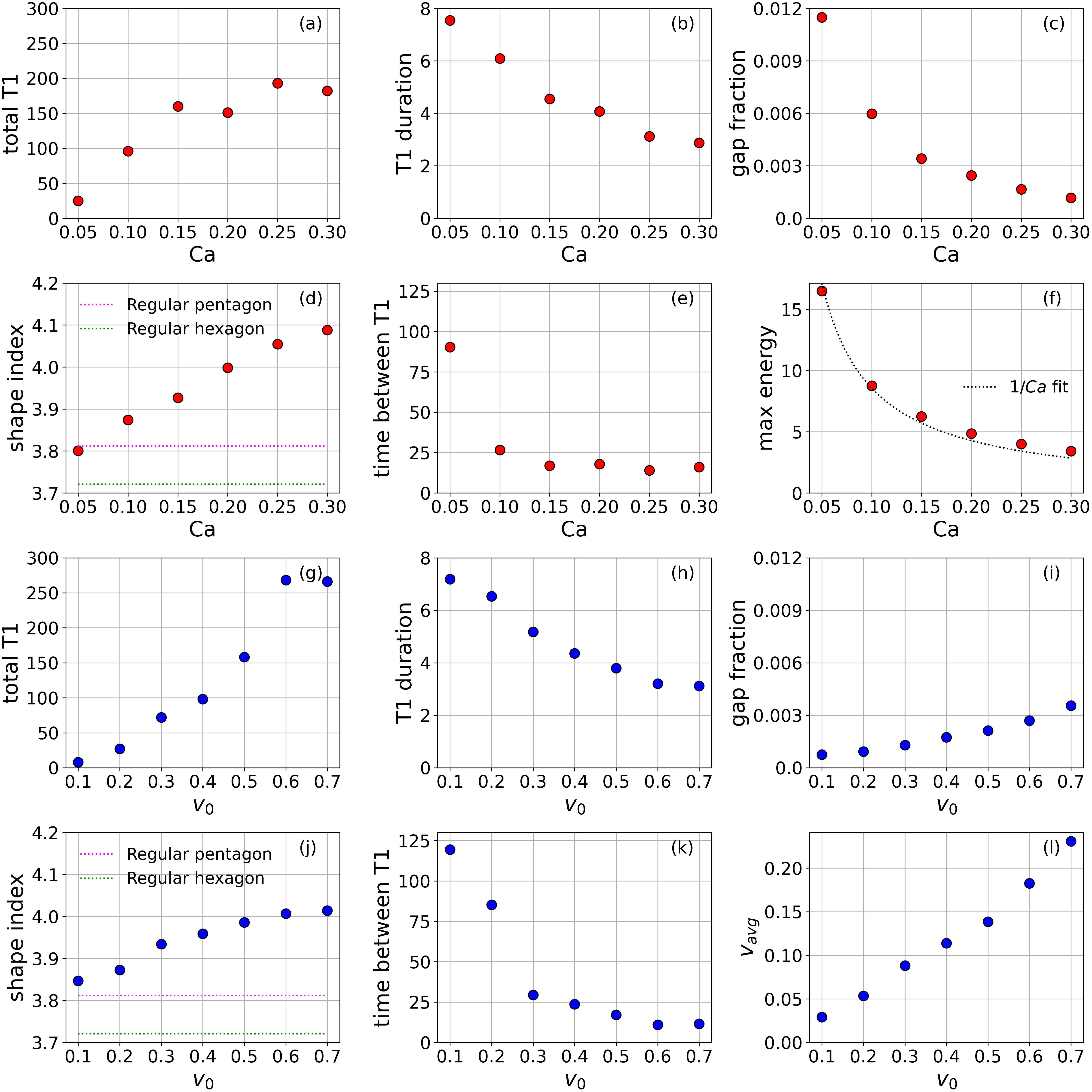}
    \caption{Dependency of various properties on deformability $Ca$ ((a) - (f)) and activity $v_0$ ((g) - (l)). \textit{Total T1}  considers the total number of T1 transitions within the considered time frame, \textit{T1 duration} is the averaged time from start to end of all T1 transitions, \textit{Gap fraction} is the extracellular space, considered as $\sum_i B(\phi_i)$ below a fixed threshold, again averaged over time, \textit{Shape index} considers the averaged shape index of the four cells involved in the T1 transitions. \textit{Time between T1} is the average time a cell spends between successive T1 transitions, \textit{Max energy} is the maximum energy reached at a T1 transition and \textit{$v_{avg}$} is the average velocity of center of mass of all cells.}
    \label{fig: time between T1 v0}
    \label{fig: time between T1 Ca}
    \label{fig: avg velocity v0}
    \label{fig: gap fraction v0}
    \label{fig: gap fraction Ca}
    \label{fig: t1 duration v0}
    \label{fig: t1 duration Ca}
    \label{fig: total t1 Ca}
    \label{fig: total t1 v0}
    \label{fig: max energy Ca}
    \label{fig: shape index Ca}
    \label{fig: shape index v0}
    \label{fig: Ca variation}
\end{figure}

%%%%%%%%%% Figure &&&&&&&&&&&

\subsection{Effect of cell deformability, activity and gaps on T1 transitions}
The asymmetric energy profile in Figure \hyperref[fig: energy prepost profile]{3a} is robust to tuning of most of the model parameters. Significant variations only occur for low values of $Ca$ and $v_0$, see Figure \hyperref[fig: energy prepost Ca]{4b, 4e}. We now analyse the effect of cell deformability and activity on T1 transitions in more detail. This requires a detailed analysis of the influence of gaps. The gap fraction is related to the confluency as $\text{confluency} = 100(1-\text{gap fraction})$. It essentially is a fixed quantity set by the initial data. We fix all parameters as per table \ref{tab: parameters} and compare two different initial cell sizes, denoted by 'low gap' with gap fraction $0.00048$ and 'high gap' with gap fraction $0.00212$. Both can be considered as confluent. The number of T1 transitions within the considered time frame is not influenced by this variation. The total numbers of T1 transitions are 162 and 158 for low and high gap cases, respectively. However, the average duration of T1 transitions is reduced by reducing the gap fraction. The values are 2.559 and 3.794 for low and high gap cases, respectively. We measure the gap fraction as the fraction of domain where $\sum_i B(\phi_i)$ is less than a fixed threshold which is set to $0.2$. This essentially excludes possible partial overlap of the diffuse interface region of cells and only accounts for gaps at tri-junctions and rosettes. This makes the measured gap fraction to depend on deformability and activity. For the considered cases low $Ca$ leads to rounder cells with stronger overlap of the diffuse interfaces of the cells, which are in contact. This leads to an increase in the measured gap fraction, see Figure \hyperref[fig: gap fraction Ca]{6c}. A similar dependency, but smaller in magnitude, is found for activity. Larger $v_0$ lead to stronger interactions between cells and thus more overlap of the diffuse interface region of cells in contact which again leads to an increase in measured gap fraction, see \hyperref[fig: gap fraction v0]{6i}. The gap fraction in both figures is the average quantity over the considered time frame. Both results and the dependencies discussed below are considered for the 'high gap' setting. 

As shown in Figure~\hyperref[fig: total t1 Ca]{6a}, the number of T1 transitions increases with increasing cell deformability parameter $Ca$. Cells that are more deformable can more easily acquire the shape deformations associated with T1 transitions. When $Ca$ is low, these deformations are energetically more expensive resulting in fewer T1 transitions. Also the duration of T1 transitions depends on $Ca$, as shown in Figure~\hyperref[fig: t1 duration Ca]{6b}. T1 transitions are shorter when cells are more deformable. We suspect that this might be due to the presence of smaller gaps at T1 transitions, as this requires less shape deformation. Figure \hyperref[fig: shape index Ca]{6d} shows the average cell shape index of the four involved cells in a T1 transition as function of cell deformability $Ca$. The shape index increases as deformability increases. The shape index of $Ca=0.05$ is less than that of a regular pentagon. The shape index of regular pentagon (3.813) was attributed as the critical shape index for jamming transition in classical vertex models \cite{Park2015} without gaps. It has been argued that gaps influence the mechanical properties and solid-liquid transition \cite{durandRelaxationTimeTopological2006}, which might explain this discrepancy, as our system is still within the fluid phase. Further details, which are related to the previous dependencies are shown in Figures \hyperref[fig: time between T1 Ca]{6e} and \hyperref[fig: max energy Ca]{6f}. Figure \hyperref[fig: time between T1 Ca]{6e} shows the average time a cell spends between successive T1 transitions as function of $Ca$. This quantity is large for low $Ca$ but decreases and plateaus to low values upon increasing $Ca$. Figure \hyperref[fig: max energy Ca]{6f} shows the maximum energy reached during a T1 transition against $Ca$. We see from the dotted curve that the maximum energy is proportional to $1 / Ca$. Recall that $1 / Ca$ scales the Cahn-Hilliard energy as per equation \eqref{eq: Cahn Hilliard Energy}. This means that $\mathcal{F}_{r_{avg}}$ is primarily affected by the Cahn-Hilliard energy, which explains the correspondence of our results with the length of T1 junctions discussed earlier and considered in \cite{durandRelaxationTimeTopological2006,curranMyosinIIControls2017}.

The dependency on $v_0$ shows qualitatively similar behaviour for the number of T1 transitions, the duration of T1 transitions, the shape index of the cells involved in T1 transitions and the time a cell spends between successive T1 transitions, see Figures \hyperref[fig: total t1 Ca]{6g, 6h, 6j, 6k}, respectively. The increase in T1 transitions and decrease in the time between T1 transitions with activity is a property of active systems, which are driven out of equilibrium. T1 transitions are topological defects and thus an indication of out of equilibrium. The decrease in duration with increasing activity can again be associated with the decrease in measured gap fraction, see \hyperref[fig: gap fraction v0]{6i}, and also the increasing shape index with activity is a direct consequence of the form of active forcing considered. Figure \hyperref[fig: total t1 Ca]{6l} shows the average velocity of center of mass of all cells as a function of v$_0$. As expected, activity is primarily converted into motion with an almost linear dependency.

\begin{figure}[htb!]
\centering
\begin{subfigure}[b]{0.32\textwidth}
\centering
\includegraphics[width=0.85 \textwidth]{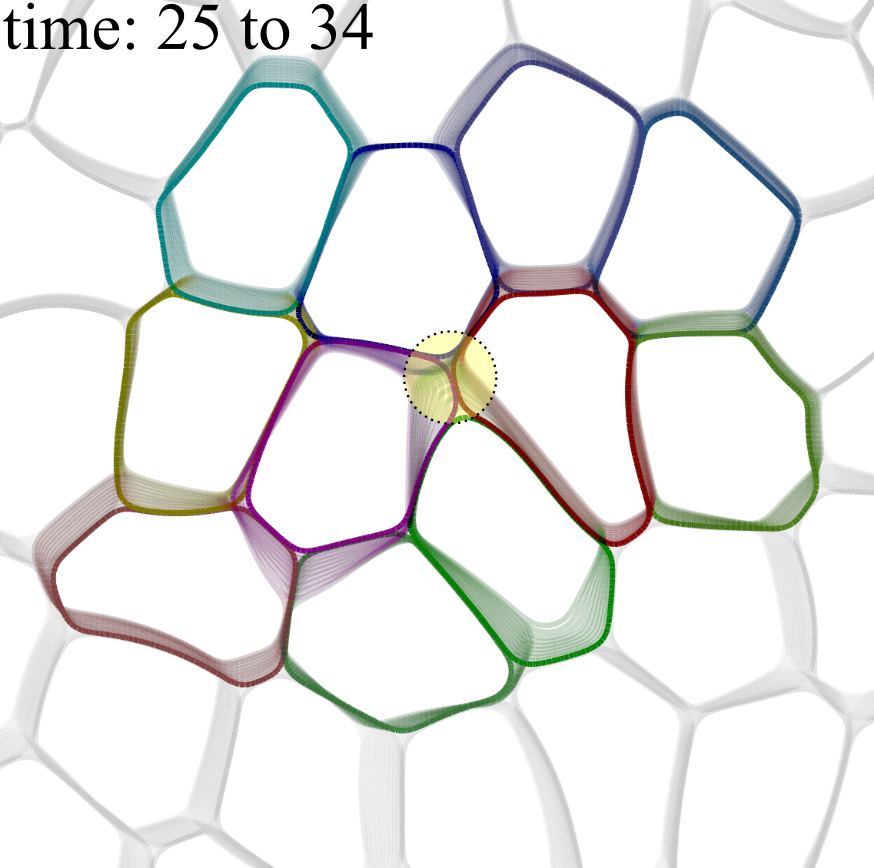}
\caption{}
\label{fig: chaining 1}
\end{subfigure}
%\hfill
\begin{subfigure}[b]{0.32\textwidth}
\centering
\includegraphics[width=0.85\textwidth]{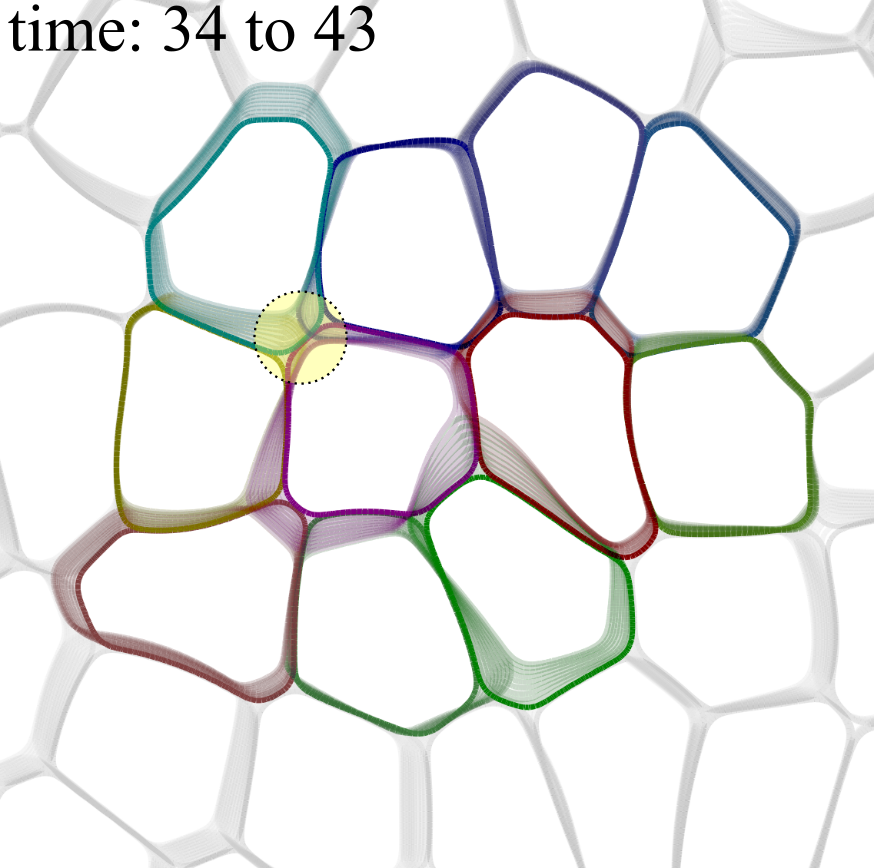}
\caption{}
\label{fig: chaining 2}
\end{subfigure}
%\hfill
\begin{subfigure}[b]{0.32\textwidth}
\centering
\includegraphics[width=0.85\textwidth]{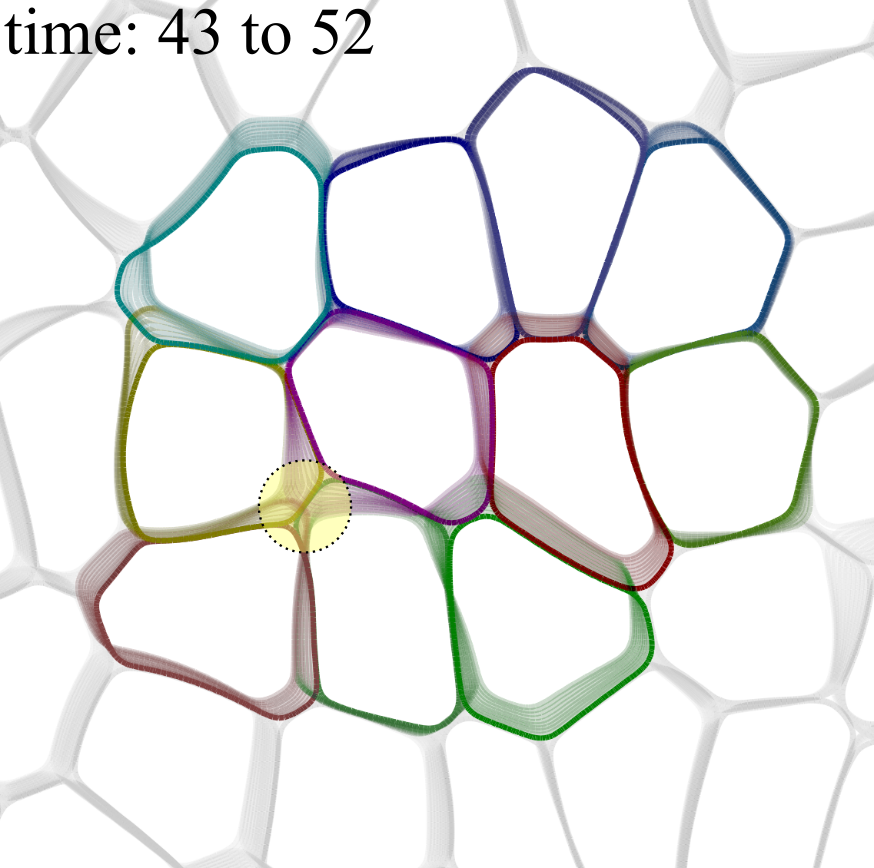}
\caption{}
\label{fig: chaining 3 }
\end{subfigure}
\begin{subfigure}[b]{0.32\textwidth}
\centering
\includegraphics[width=0.85\textwidth]{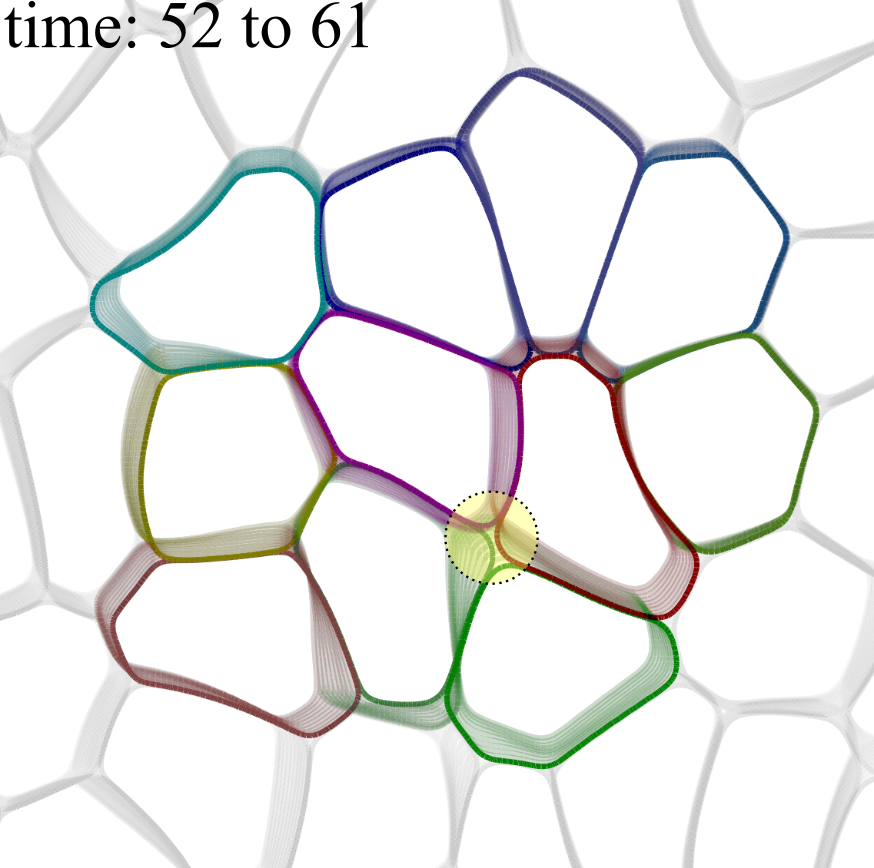}
\caption{}
\label{fig: chaining 4}
\end{subfigure}
\begin{subfigure}[b]{0.32\textwidth}
\centering
\includegraphics[width=0.85\textwidth]{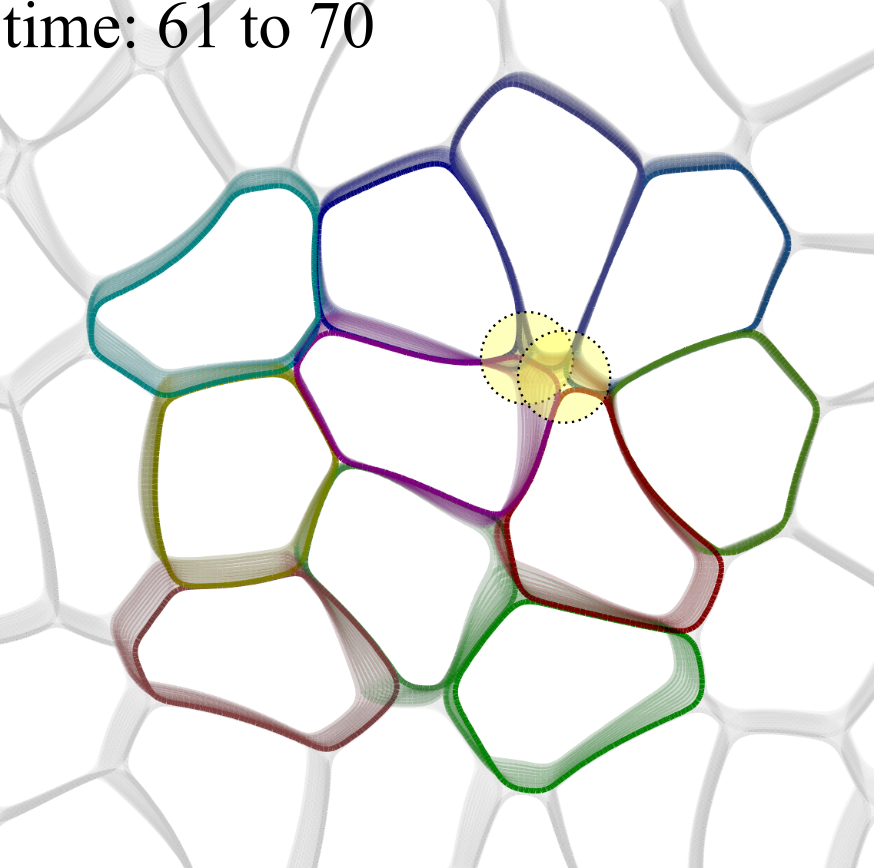}
\caption{}
\label{fig: chaining 5}
\end{subfigure}
\begin{subfigure}[b]{0.32\textwidth}
\centering
\includegraphics[width=0.85\textwidth]{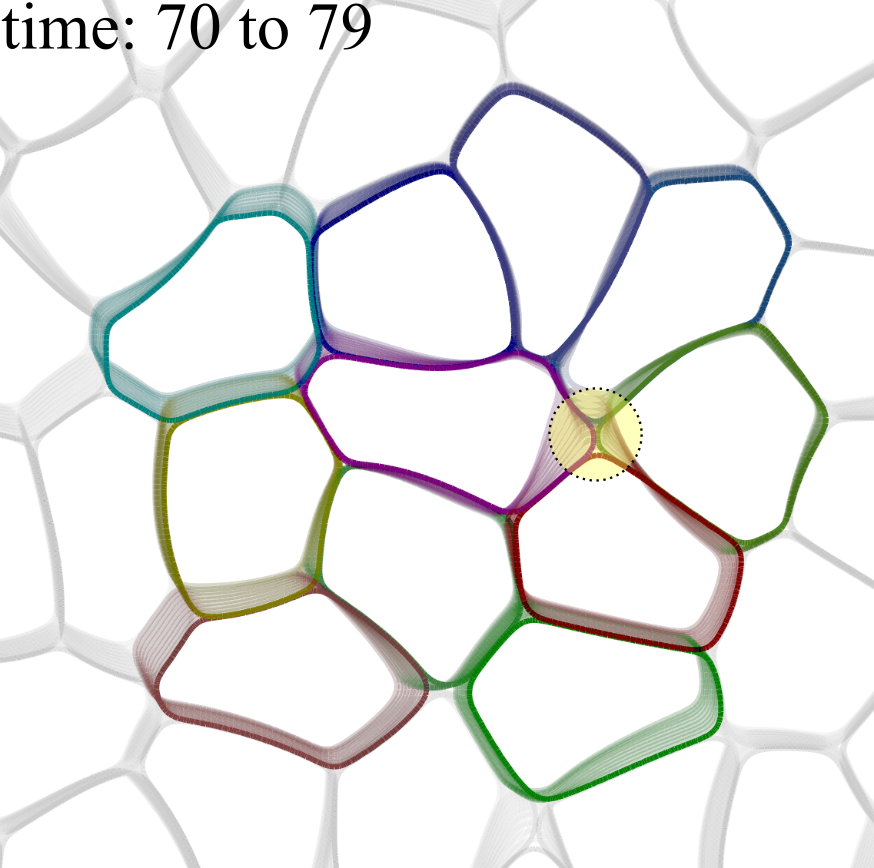}
\caption{}
\label{fig: chaining 6}
\end{subfigure}
\caption{Chaining of T1 transitions. Each panel is a montage of $10$ snapshots of tissue configurations taken successively at constant times intervals. Latest time is represented by the cell shapes marked in the darkest color shades. The cores of the T1 transitions are highlighted in yellow. Also see Supplementary Movie 3.}
\label{fig: chaining of T1 transitions}
\end{figure}

%%%%%%%%%%&&&&&&&&&&&
\begin{figure}[htb!]
\centering
\includegraphics[width=0.9\textwidth]{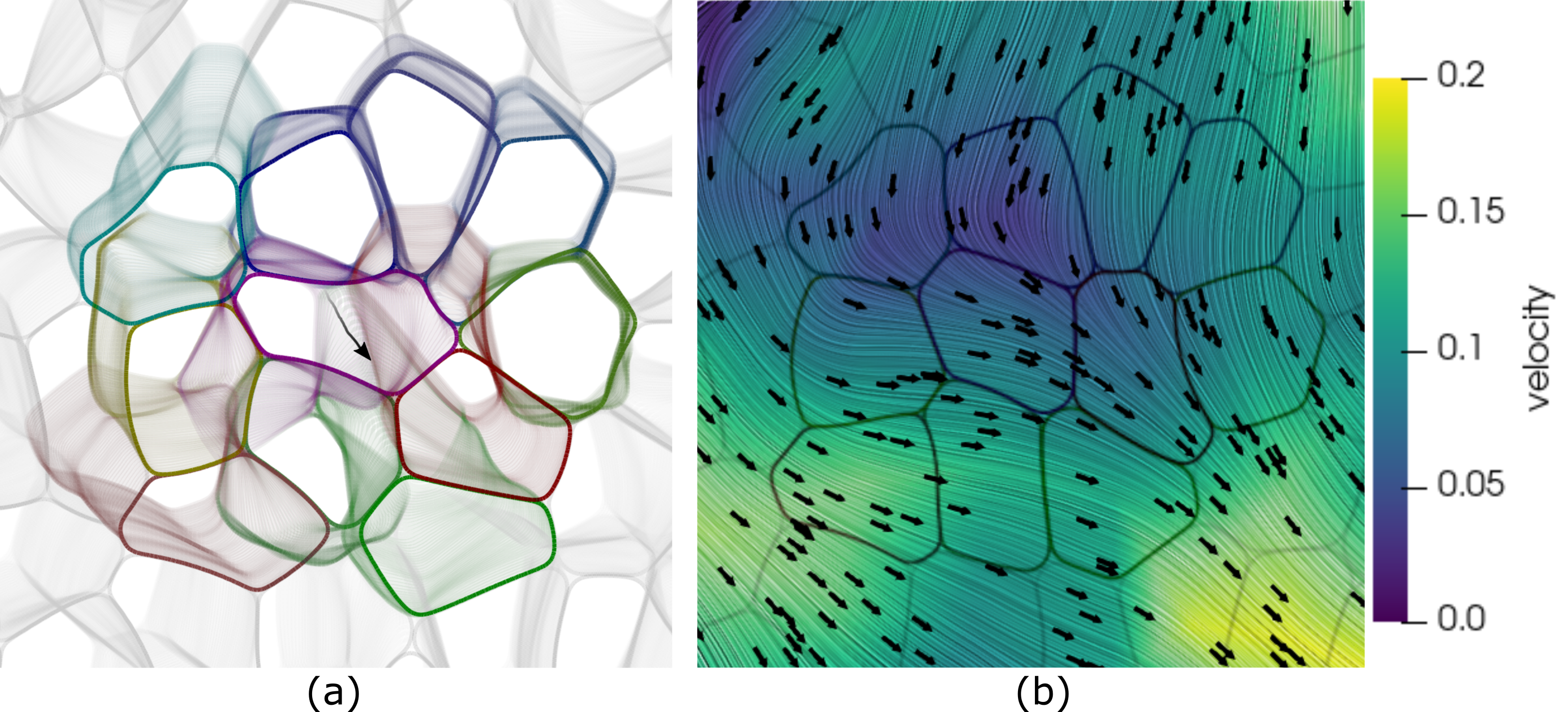}
\caption{(a) Montage of tissue snapshots from time $t=25$ to $t=79$ (see figure \ref{fig: chaining of T1 transitions}). The black path is the trajectory of the center of mass of the 11 coloured cells. (b) LIC visualization of streamlines, magnitude (color) and direction (black arrows) of the flow velocity. The velocity and the cell boundaries correspond to time $t=52$.}
\label{fig: deformations due to chaining of T1 transitions}
\end{figure}

\subsection{Chaining of T1 transitions} \label{Sec: T1 Chain}
So far, we have analysed robust statistical properties of T1 transitions within their cores. However, we have also seen that these local features influence the position and shape of the four cells involved in a T1 transition, and their neighbours. This can induce new T1 transitions and lead to the formation of 
chains of T1 transitions as illustrated in Figure~\ref{fig: chaining of T1 transitions}. Each of these images consists of $10$ tissue states captured at equally-spaced time instants and overlaid on top of each other. The cell shapes outlined in the darkest colors correspond to the latest time. The yellow circles mark the cores of the T1 transitions at those time instants. %The cartoon arrows are meant to depict the general direction of motion of the cells.
The chaining of T1 transitions is a result of the assumptions on constant cell area and a confluent tissue. Any cell deformation associated with a T1 transition induces deformation of the neighbouring cells and thereby increases the possibility of new T1 transitions. This is further enhanced by activity and the considered propulsion mechanism which favours the direction of elongation.  
%In our model, the cells have fixed area. When a cell moves to a previously empty region, it leaves behind an equally large empty region. The neighbours of the cell rush in to fill this new empty region resulting in a chain of movements. In some instances, a neighbour might not fill a region due to being restricted by cells around it. In such situations, cell-cell contacts can be broken triggering T1 transitions. In other instances, multiple cells can rush into the empty space left behind by a single cell. The multiple cells would then each leave behind less empty space compared to the single cell. This means, the effect of the T1 transition would reduce down the chain as the motion is distributed among many cells. In essence, a T1 transition can trigger a chain of T1 transitions.

This chaining of T1 transitions is also observed experimentally in sheared foams~\cite{rosaNucleationGlideDislocations1998} and in our simulations of passive foams which are sheared with a constant shear velocity profile. For $v_0 =0$, typically one or two T1 transitions occur due to the initial non-equilibrium configuration of the tissue. As cells relax toward an equilibrium state, their motility is reduced which prevents any further T1 transitions. The situation for small $v_0$ is similar. The tissue becomes jammed by cells being caged amongst their neighbours and no T1 transitions occur \cite{wenzelMultiphaseFieldModels2021}. Furthermore, when cell deformability ($Ca$) is low, the energetic cost for cell deformations that are necessary to undergo T1 transitions is high, which prevents or at least reduces T1 transitions and the tissue also becomes jammed \cite{wenzelMultiphaseFieldModels2021}. This corresponds to the low number of T1 transitions in Figure~\hyperref[fig: energy prepost Ca]{4b, 4e} for low $Ca$ and low $v_0$, respectively.

However, in the considered case in Figure~\ref{fig: chaining of T1 transitions} we are far away from jamming and the chaining of T1 transitions leads to cell deformation propagating to larger scales. This is highlighted in Figure~\hyperref[fig: chaining of T1 transitions]{8a}, which shows the evolution of the cell tissue in the whole time window considered in Figure~\ref{fig: chaining of T1 transitions} together with the trajectory of the center of mass of the colored cells, which highlights the movement on larger spatial scales. The chaining of T1 transitions is also a source of large-scale flows as evidenced in Figure~\hyperref[fig: chaining of T1 transitions]{8b}.
We consider the velocities of the centers of mass of all cells, average this quantity with the neighboring cells and  construct a continuous velocity field by interpolating in space. The velocity field is shown together with the cell boundaries at $t = 52$. The mean direction corresponds with the direction of the black path shown in Figure~\hyperref[fig: chaining of T1 transitions]{8a}. However, as the variations in magnitude and direction of the flow field in Figure~\hyperref[fig: chaining of T1 transitions]{8b} indicate, T1 transitions can also induce fluctuations and could play an important role in sustaining chaotic flows (active turbulence) in cell tissues~\cite{Doostmohammadi2018,wenzelDefectsActiveNematics2021,alertActiveTurbulence2022}.
%---------------------
\section{Discussion} \label{Sec:Discussion}

Large-scale tissue deformation requires cellular rearrangements. The simplest rearrangement in confluent cell tissue is a T1 transition. We have analysed these neighbour exchanges among cells in detail using a multi-phase field model and identified a characteristic asymmetric energy profile, see Figure \ref{fig: energy prepost profile}. The energy profile has a peak at the T1 transition. The profile is asymmetric with a strong increase in energy before the T1 transition and a sudden decrease after the T1 transition which is followed by a slow relaxation. Detailed studies on the dependency of this profile on model parameters show robustness to variations in most parameters. They also allowed to associate the strong energy increase before the T1 transition with the strength in activity. This region is characterized by an accumulation of energy to reach the energy barrier at the T1 transition. This is achieved by probing several possibilities of direction of movement and shape deformation. This process is enhanced by activity, which is quantified by Figure \hyperref[fig: energy prepost velocity]{4e}. In contrast to this the sudden relaxation after the T1 transition can clearly be associated with energy relaxation. It is almost independent of activity, see Figure \hyperref[fig: energy prepost velocity]{4e}, and cell deformability, see Figure \hyperref[fig: energy prepost velocity]{4b}, and also present in sheared foams, see Figure \hyperref[fig: energy prepost velocity]{4f}. We would like to remark that the behaviour is independent but the actual slope and duration of this regime depends on deformability, as the energy is scaled in Figure \hyperref[fig: energy prepost velocity]{4b}. The sudden decrease is associated with a
steep gradient in the energy landscape in one direction set by the deformation of the cells in the core of the T1 transition. The third characteristic region, the slow relaxation, depends on activity and cell deformability. This relaxation profile provides insight in the mechanical properties of the tissue. Similar energy profiles have been obtained by actuation and relaxation of magnetic microdroplets which are injected into the tissue \cite{mongeraFluidtosolidJammingTransition2018,kimEmbryonicTissuesActive2021,durandRelaxationTimeTopological2006}. In these experiments a slow relaxation is associated with the fluidization of the tissue \cite{mongeraFluidtosolidJammingTransition2018,kimEmbryonicTissuesActive2021}, while stagnation of the relaxation indicates more solid-like behaviour \cite{durandRelaxationTimeTopological2006} and is associated with irreversible (plastic) tissue rearrangements. We postulate that these mechanical characterizations can also be obtained from the energy decay of the T1 transitions.

In the considered confluent tissue the type of interaction between the cells, if repulsive or repulsive and attractive, seems to play a minor role on the characteristic energy profile of a T1 transition, see Figure 
\hyperref[fig: energy prepost velocity]{4c}. However, the degree of confluency is known to influence the solid-fluid phase transition \cite{mongeraFluidtosolidJammingTransition2018}. Increasing the extracellular space enhances fluidization. While we only consider the fluid phase, we observe an increased duration of T1 transitions for larger extracellular space. 
A finite duration of T1 transitions in cell tissues has been associated with molecular processes and is considered in an adhoc manner in vertex models \cite{erdemci-tandoganEffectCellularRearrangement2021}. Within the multi-phase field model a finite duration is a result of the mechanical properties of the cells and the their interactions. An increased duration of T1 transitions is observed for low deformability and low activity, see Figures \hyperref[fig: t1 duration Ca]{6b} and \hyperref[fig: t1 duration v0]{6h}, respectively. Both indicating more solid-like behaviour, which is consistent with \cite{erdemci-tandoganEffectCellularRearrangement2021}, where increased duration of T1 transitions leads to decreasing the overall number of T1 transitions and a possible stiffening of the global tissue mechanics. However, these results don't take extracellular space into account. 

Even if characterized locally, due to the confluent cell tissue, large enough deformations induced by T1 transitions lead to permanent cell deformations in the 
neighbourhood, which can trigger other T1 transitions, leading to a chaining effect. This behaviour is associated with the foam-like architecture and consistent with previously reported nonlinear tissue mechanics \cite{mongeraFluidtosolidJammingTransition2018}. It is this chaining of T1 transitions which allows for large-scale tissue deformations and flow patterns which can be associated with sustaining chaotic flows, see Figure~\hyperref[fig: chaining of T1 transitions]{8b}. 

We believe these results also to hold in more general situations, e.g. for varying cell sizes and varying mechanical cell properties.

\section{Numerical Methods} \label{Sec:Methods}

%---------------------------
\subsection*{Model Parameters}
    Unless otherwise specified, we use the model parameters as per Table \ref{tab: parameters}
    %%%%%%%%%%% Table of parameters %%%%%%%%%%%%%
    \begin{table}[htb!]
    \centering
    \begin{tabular}{|c|c|c|c|c|c|c|c|c|c|c|}
        \hline
        $\tau$ & $\tau_{save}$ & $T$ & $L$ & $\epsilon$ & $\boldsymbol{v_0}$ &  $a$ & $Ca$ & $In$ & $D_r$ & $\alpha$ \\
        \hline 
       0.005 & 0.5 & 150 & 100 & 0.15 & 0.5 & 1.5 & 0.2 & 0.1 & 0.1 & 0.1 \\
        \hline 
    \end{tabular}
    \caption{Default values of the model parameters.}
    \label{tab: parameters}
    \end{table}
    %%%%%%%%%%%%%%%%%%%%

\subsection*{Finite element simulations}

The simulations are run for time interval $[0,T]$ discretised into $N_t$ units with a uniform timestep size $\tau$, i.e. $T = N_t\tau$. We employ a semi-implicit discretization in time. Discretization in space follows the finite element method. We adaptively refine the diffuse interface and employ a parallelization approach which scales with the number of cells. For details we refer to \cite{Marth2016,Marth2016a,praetoriusCollectiveCellBehaviour2018,Wenzel2019,wenzelMultiphaseFieldModels2021,jainImpactContactInhibition2022}. The algorithm is implemented in the open-source library AMDiS \cite{Vey2007,Witkowski2015}.

\subsection*{Detecting T1 transitions}\label{Sec: Detecting T1s}
The T1 transitions are detected by tracking the neighbour relations of all cells. If two cells $A$ and $B$ are in contact, their neighbour relation is denoted by $(A, B)$ or $(B, A)$, both of which are equivalent. Suppose, there are four cells as in the Figure~\ref{fig: T1 transition}. The set of neighbour relations between these four cells before, during and after a T1 transition are $\{(A, B), (B, C), (C, D), (D, A), (B, D)\}$, $\{(A, B), (B, C), (C, D), (D, A)\}$ and $\{(A, B), (B, C), (C, D), (D, A), (A, C)\}$ respectively. Before and after a T1 transition, there are 5 distinct neighbour relations between the four cells. The sets of relations before and after a T1 transition have four elements in common. These common elements make up the set of relations during a T1 transition. The duration of a T1 transition is time difference when the number of neighbour relations between the four cells change from 5 to 4 and back to 5. 
\subsection*{Sensitivity of $f_{r_{avg}}$ on ${r_{avg}}$}\label{Sec: Sensitivity to coarse graining}
The coarse graining region of a point $\mathbf{p}$ is the region with all points $\mathbf{x}$ such that $|\mathbf{p}-\mathbf{x}| < r_{avg}$. As the free energy is high at the cell edges, the points which include the edges within its coarse graining region around it would have high $f_{r_{avg}}$. Moreover, points with triple junctions (where 3 edges meet) within its coarse graining region would have a higher $f_{r_{avg}}$ due to the presence of longer total length of cell edges. Usually at a given time, $f_{r_{avg}}$ has peaks near the T1 epicenter. This is because, the region around it would have either two triple junctions along with a gap as seen in the snapshots of Figure~\ref{fig: T1 transition}. Also, it is clear that points that do not have any cell edges within its coarse graining region, would have zero $f_{r_{avg}}$. We have found that increasing the $r_{avg}$ loses information about the T1 transition in the value of $f_{r_{avg}}$ at the epicenter. A larger coarse graining region would entail a larger contribution from the bulk of the interior of the cell and would reduce $f_{r_{avg}}$ at the epicenters such that $f_{r_{avg}}$ at epicenters would not be uniquely discerned as a signature of a T1 transition. On the other hand, reducing $r_{avg}$ would mean that we might not encompass the information of the two triple junctions and the gap formed during the T1 transition. It also increases the deviations in the statistics that we describe. Moreover, if the energy along the length of the edge is uniform then the energy field $f_{r_{avg}}$ at a point gives an approximate measure of length of edges within the coarse graining region around that point.

\section*{Data availability}
All data are available from the corresponding author upon reasonable request. The AMDiS implementation and additional codes for pre- and postprocessing are available from the corresponding
author upon reasonable request

\section*{Supplementary Information}
Supplementary Movie 1
\noindent
\\
Supplementary Movie 2
\noindent
\\
Supplementary Movie 3

\bibliography{t1_transitions}

\section*{Acknowledgements}

This project has received funding from the European Union's Horizon 2020 research and innovation programme under the Marie Skłodowska-Curie grant agreement No 945371. We acknowledge computing resources provided within project WIR at ZIH at TU Dresden.

\section*{Author contributions statements}

H.P.J. implemented the codes, performed all simulations, analysed data and contributed to conceptual development and manuscript writing. A.V. and L.A. contributed to supervision, conceptual development, data analysis and manuscript writing. 

\section*{Additional information}
\textbf{Competing interests} 
The authors declare no competing interests.

\end{document}